\documentclass{jfm}
\renewcommand{\figurename}{figure}
\usepackage[utf8x]{inputenc}
\usepackage{psfrag}
\usepackage{amssymb, amsmath}
\usepackage{siunitx}
\usepackage{color}

\usepackage{url}

\DeclareSIUnit{\wtpercent}{wt\%}

\usepackage{upgreek}
\begin{document}

\shorttitle{Evaporating Ouzo Droplets} 
\shortauthor{C. Diddens et al.} 

\title{Evaporating pure, binary \& ternary droplets: thermal effects \& axial symmetry breaking}

\author{
Christian~Diddens\corresp{\email{c.diddens@utwente.nl}}\aff{1,2},
Huanshu~Tan\aff{1},
Pengyu~Lv\aff{1},
Michel~Versluis\aff{1},
J.~G.~M.~Kuerten\aff{2,3},
Xuehua~Zhang\aff{4,1},
Detlef~Lohse\corresp{\email{d.lohse@utwente.nl}}\aff{1,5}
}

\affiliation{
\aff{1}
Physics of Fluids group, Department of Science and Technology, Mesa+ Institute, and 
J.~M.~Burgers Centre for Fluid Dynamics, University of Twente, P.O. Box 217, 7500 AE Enschede, The Netherlands
\aff{2}
Department of Mechanical Engineering, Eindhoven University of Technology, P.O. Box 513, 5600 MB Eindhoven, The Netherlands
\aff{3}
Faculty EEMCS, University of Twente, P.O. Box 217, 7500 AE Enschede, The Netherlands
\aff{4}
Soft Matter \& Interfaces Group, School of Engineering, RMIT University, Melbourne, VIC 3001, Australia
\aff{5}
Max Planck Institute for Dynamics and Self-Organization, 37077 G\"ottingen, Germany
}

\maketitle

\begin{abstract}
The Greek aperitif Ouzo is not only famous for its specific anise-flavored taste, but also for its ability to turn from a transparent miscible liquid to a milky-white colored emulsion when water is added. Recently, it has been shown that this so-called Ouzo effect, i.e. the spontaneous emulsification of oil microdroplets, can also be triggered by the preferential evaporation of ethanol in an evaporating sessile Ouzo drop, leading to an amazingly rich drying process with multiple phase transitions [H. Tan et al., Proc. Natl. Acad. Sci. USA \textbf{113}(31) (2016) 8642]. Due to the enhanced evaporation near the contact line, the nucleation of oil droplets starts at the rim which results in an oil ring encircling the drop. Furthermore, the oil droplets are advected through the Ouzo drop by a fast solutal Marangoni flow.

In this article, we investigate the evaporation of mixture droplets in more detail, by successively increasing the mixture complexity from pure water over a binary water-ethanol mixture to the ternary Ouzo mixture (water, ethanol and anise oil). In particular, axisymmetric and full three-dimensional finite element method simulations have been performed on these droplets to discuss thermal effects and the complicated flow in the droplet driven by an interplay of preferential evaporation, evaporative cooling and solutal and thermal Marangoni flow. By using image analysis techniques and micro-PIV measurements, we are able to compare the numerically predicted volume evolutions and velocity fields with experimental data. The Ouzo droplet is furthermore investigated by confocal microscopy. It is shown that the oil ring predominantly emerges due to coalescence.

\end{abstract}
\section{Introduction}
\label{sec:intro:intro}
Evaporation of sessile droplets is a ubiquitous phenomenon which is utilized in a wide-spread range of applications, e.g. inkjet printing, coating and spray cooling. 
The pioneering work of \citet{Deegan1997a}, explaining the so-called coffee-stain effect, incented the scientific investigation of evaporating droplets in general.

While the evaporation process of droplets consisting of a pure liquid is mainly understood \citep{Cazabat2010a}, multi-component droplets show in general far more complex evolutions during evaporation. This is due to the complicated coupling of multi-component evaporation, flow of the mixture and possibly also thermal effects. Even for binary mixtures, the presence of the second component can result in non-monotonic contact angle evolutions \citep{Rowan2000a,Sefiane2003a,Cheng2006a,Wang2008a,Shi2009a}, initial condensation of the less volatile component \citep{Innocenzi2008a} or entrapped residuals of the more volatile component at later times \citep{Sefiane2008a,Liu2008a,Diddens2017a,Diddens2017b}. Furthermore, the flow in the droplet is primarily driven by the solutal Marangoni effect. In particular, it has been shown that water-ethanol droplets exhibit initially chaotic and highly non-axisymmetric flows, followed by a fast transition to nearly axisymmetric flow and outward radial flow towards the end of the lifetime \citep{Christy2011a,Bennacer2014a,Zhong2016a}. By adding surfactants and surface-absorbed polymers, the Marangoni flow can be controlled, leading to homogeneous deposition patterns \citep{Kim2016a}. Neighboring binary mixture droplets can also interact through the vapor phase which allows for the assembly of intriguing autonomous fluidic machines \citep{Cira2015a}. Also the dissolution of a binary droplet in a third liquid shows highly nontrivial and unexpected behavior \citep{Chu2016a,Dietrich2016b}.

Recently, it has been shown that ternary mixture droplets, as the next more general liquid, can show an even richer evolution process: By investigating the evaporation of an Ouzo drop (ethanol, water and anise oil), we have revealed multiple phase transitions, where the drop temporarily changes its appearance from an initially transparent liquid to a milky-white colored emulsion \citep{Tan2016a}. The reason lies in the so-called Ouzo effect \citep{Vitale2003a}, i.e. the spontaneous emulsification of oil microdroplets once the local ethanol concentration has reduced by preferential evaporation below a specific threshold. Since the evaporation rate for droplets with contact angles below $\SI{90}{\degree}$ is highest at the contact line, the onset of the Ouzo effect starts near the rim which additionally results in an oil ring encircling the drop. While most of the oil droplets are quickly transported by solutal Marangoni flow through the entire droplet, also sessile oil droplets immersed in the surrounding drop were observed -- so-called surface nanodroplets \citep{Zhang2007a,Lohse2015a,Zhang2015b}. If the contact angle of an Ouzo droplet exceeds $\SI{90}{\degree}$, the Ouzo effect initially occurs close to the apex, i.e. where the evaporation is pronounced in this case, and no persistent oil ring formation can be observed \citep{Tan2017a}. Instead, a continuous oil phase wraps up the drop due to Marangoni forces.

Although the numerical results for the volume evolution of an Ouzo droplet are in good agreement with the experimental data, there are some drawbacks in the study of \citet{Tan2016a}: On the one hand, the applied lubrication theory is only valid for very flat droplets, but the experimentally found contact angle of the Ouzo droplet temporarily exceeds $\SI{80}{\degree}$, which causes a considerable inaccuracy of the predicted flow velocity \citep{Diddens2017a,Diddens2017b}. On the other hand, it is well known from the afore-mentioned experiments that the flow in ethanol-water droplets is initially highly non-axisymmetric, which cannot be covered by the used axisymmetric model. Finally, the influence of the latent heat of evaporation has been neglected and the relative humidity had to be adjusted in order to match the experimental volume evolution. In particular on a thin substrate, as used in the experiment, thermal effects can play a crucial role \citep{Diddens2017b,Tan2017a}.

In this study, we take advantage of a finite element method model to investigate the flow velocity inside evaporating multi-component droplets in more detail, i.e. without being subject to the mentioned limitations of our previous study. We start with a pure water droplet as the simplest case and successively generalize it to a binary water-ethanol droplet and to the ternary Ouzo droplet. This procedure allows us to discuss possible disagreements between simulations and experiments for complicated mixtures in the light of the agreement for simpler mixtures. Furthermore, we present the to our knowledge first full three-dimensional model of evaporating sessile multi-component droplets. This model comprises multi-component evaporation, evaporative cooling, thermal and solutal Marangoni flow as well as composition- and temperature-dependent fluid properties. Within the framework of this detailed model, we are able to investigate the afore-mentioned axial symmetry breaking of the flow and compare it to micro-PIV measurements and a qualitative analysis by confocal microscopy.  

Details about the performed experiments are described in section \ref{sec:exp:exp}. In section \ref{sec:axisymm:axisymm}, the axisymmetric finite element method model is outlined and its results are compared with the corresponding experimental data. The model is generalized to three dimensions in section \ref{sec:threedim:threedim} to account for non-axisymmetric flow and the numerical results are compared with experimental micro-PIV measurements. Finally, in section \ref{sec:confocal:confocal}, we investigate the convection of the oil-microdroplets in the ternary Ouzo droplet and their behavior at the contact line with the aid of confocal microscopy. We reveal how the oil ring encircling the Ouzo droplet is primarily emerging due to coalescence of oil droplets at the contact line. 

\section{Experimental Setup}
\label{sec:exp:exp}

\subsection{Droplet Composition}
\label{sec:exp:composition}
In total, we have experimentally investigated the evaporation of three different types of droplets, namely a pure water droplet, a binary water-ethanol droplet and a ternary Ouzo droplet.
The liquids used for these droplets were pure Milli-Q water [produced by a Reference A+ system (Merck Millipore) at \SI{18.2}{\mega\ohm{\cdot}\centi\meter} (at \SI{25}{\degreeCelsius})], a mixture of $\SI{37.88}{\percent}$ (wt/wt) Milli-Q water and $\SI{62.12}{\percent}$ (wt/wt) ethanol [Sigma-Aldrich; $\geq$ \SI{99.8}{\percent}], and a mixture of $\SI{37.24}{\percent}$ (wt/wt) Milli-Q water, $\SI{61.06}{\percent}$ (wt/wt) ethanol and $\SI{1.70}{\percent}$ anise oil [Sigma-Aldrich], respectively.
Furthermore, for the experimental micro-PIV measurements described later on in section \ref{sec:exp:micropiv}, tracer particles have been added to the binary water-ethanol droplet.

\subsection{Setup and Image Analysis}
\label{sec:exp:imageana}
The droplets were deposited through a teflonized needle [Hamilton; 8646-02] by a motorized syringe pump [Harvard; PHD 2000] on a flat hydrophobic octadecyltrichlorosilane (OTS)-glass surface with a thickness of $d_\text{s}=\SI{0.17}{\milli\meter}$.
The advancing and receding contact angles of water on the surface are $\ang{112}$ and $\ang{98}$, respectively. 
The entire evaporation process of the droplets was recorded by a CCD camera [Ximea; MD061MU-SY, 3 frames per second (fps) at $\num{1372} \times \num{1100}$ pixel resolution] equipped with a long-distance microscope system [Infinity; Model K2 DistaMax] for side-view recordings and a digital SLR camera [Nikon D750] equipped with a CMOS sensor [24 frames per seconds (fps) at $\num{1920} \times \num{1080}$ pixels resolution] attached to a high-magnification zoom lens system [Thorlabs; MVL12X3Z] for top-view recordings.
The side-view illumination was provided by a homemade collimated LED source.
A universal hand-held probe [Omega; HH-USD-RP1, accuracy relative humidity is $\pm$ \SI{2}{\percent} over 10 to \SI{90}{\percent} @\SI{25}{\degreeCelsius} and a temperature accuracy  of $\pm$ \SI{0.3}{\kelvin} @\SI{25}{\degreeCelsius}] was used to measure the relative humidity and temperature in the laboratory at a sampling rate of one per second. 
A similar sketch of the setup is described in detail in \citet{Tan2016a}.

The image analysis was performed by a custom-made MATLAB program, through which all of the geometric parameters at every frame were successfully determined.
For the millimetric droplets in this study, the characteristic lengths of the droplets are smaller than the capillary length scale, which is equal to $\SI{2.7}{\milli\meter}$ for water and $\SI{1.7}{\milli\meter}$ for ethanol. Thus, gravity effects influencing the shape of the droplets can be disregarded. 
For pure water droplets and binary water-ethanol droplets, a spherical-cap shape assumption was used over the whole evaporation process. The program fits a circle to the contour of the droplet silhouette in side view. The contact angle $\theta$ was calculated based on the position of the intersection of the base line and the fitted circles.
Since Ouzo droplets temporarily show a very characteristic deviation from the spherical cap shape due to the appearance of an oil ring at the rim, only the top part of the contour above the oil ring was fitted by a circle. A polynomial fit was applied to the profile of the oil ring and gave the contact angle $\theta$. The contact angle $\theta^*$ of the spherical cap shaped top part was calculated based on the position of the intersection point where the fitted line of the oil ring profile and the fitted circle of the top contour cross. The droplet volume $V$ was calculated by integrating the volumes of horizontal disk layers and assuming rotational symmetry of each layer with respect to the vertical axis. 

\subsection{Micro Particle Image Velocimetry (micro-PIV)}
\label{sec:exp:micropiv}

\begin{figure}\centering\includegraphics[width=0.9\textwidth]{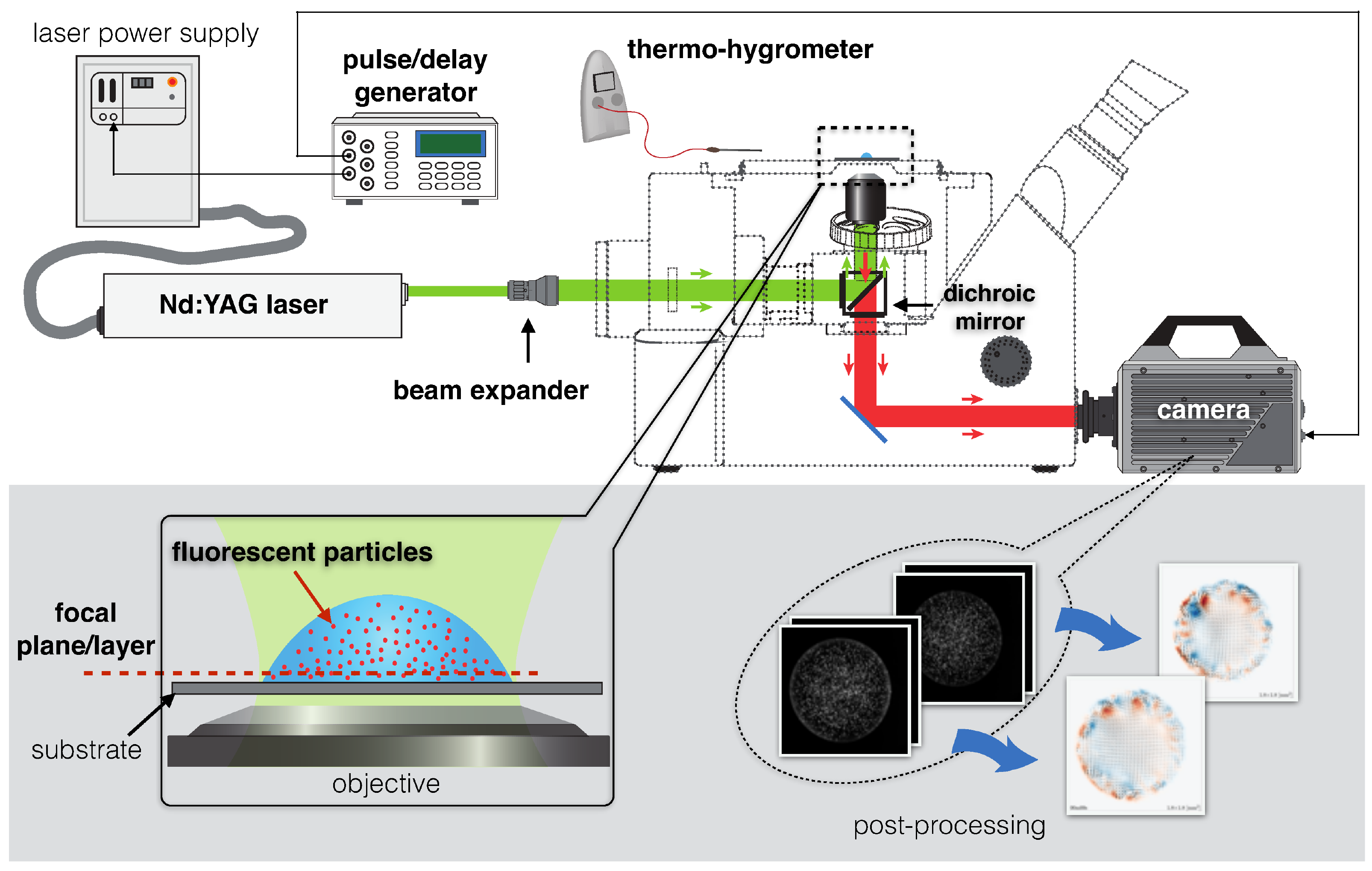}
\caption{Experimental setup of the micro-PIV system. The dual-cavity laser and the high-speed camera are synchronized to record consecutive image pairs. To apply fluorescence techniques, the inverted microscope is equipped with a dichroic cube and the droplet is seeded with fluorescent particles. The focal plane of the microscope is placed at the bottom of the droplet, i.e. close and parallel to the substrate. }
\label{fig:exp:pivscheme}
\end{figure}

To determine the velocity in the binary water-ethanol droplet, Micro Particle Image Velocimetry (micro-PIV) measurements were performed.
To that end, the binary water-ethanol mixture was seeded with fluorescent particles. Per $\SI{1}{\milli\liter}$ of the water-ethanol solution, $\SI{30}{\micro\liter}$ of aqueous tracer particle solution [microParticles GmbH; PS-FluoRed-5.0: Ex/Em $\SI{530}{\nano\meter}$/$\SI{607}{\nano\meter}$] was used. The diameter of these tracer particles is $d_{\text{p}}=\SI{5}{\micro\meter}$. 

An overview of the experimental micro-PIV setup is given in \figurename~\ref{fig:exp:pivscheme}. It consists of a dual-cavity Nd:YAG laser [Litron; NANO S 65-15PIV], a high-speed camera [Photron; Fastcam SA-X2 64 GB] and an inverted microscope [Olympus; GX-51] which is equipped with a dichroic cube.
The laser and the high-speed camera are synchronized by a pulse/delay generator [BNC; Model-575].
The laser beam illuminates the droplet from below through a 10$\times$ magnification objective lens with a numerical aperture (NA) of 0.30 (zoom-in of \figurename~\ref{fig:exp:pivscheme}).
The objective focal plane was placed just above the surface at approximately $\SI{13}{\micro\meter}$ to match the thickness of the focal plane, or the depth of field (DOF), which was calculated to be $\SI{13.4}{\micro\meter}$. Thus, we had a sharp image of the tracer particles closest to the substrate.

In this way, 10 consecutive image pairs with an interframing time of $\SI{4}{\milli\second}$ were taken per second.
To calculate the velocity field, the obtained images were first post-processed with a custom MATLAB code to enhance the contrast.
Then, the image pairs were analyzed with PIVlab \citep{thielicke2014pivlab, williamPHD}, using an interrogation window of 64 $\times$ 64 pixels, corresponding to $\SI{128}{\micro\meter} \times \SI{128}{\micro\meter}$. An interrogation window overlap of $\SI{75}{\percent}$ leads to a $\SI{32}{\micro\meter}$ vector spacing. 

To qualify the degree to which the tracer particles exactly follow the flow, we analyzed the Stokes number and the ratio of Stokes number to a buoyancy-corrected Froude number \citep{varghese2016}. The Stokes number is defined as $\mbox{\it St} \equiv t_0 u_{\text{max}}/ r_{\text{c}}$, where $t_0 \equiv  \frac{\rho_{\text{p}}d_{\text{p}}^2}{18 \mu_{\text{f}}}(1+\rho_{\text{f}}/(2\rho_{\text{p}}))$ is the characteristic time, $u_\text{max}$ the maximum fluid velocity ($\sim \SI{10}{mm.s^{-1}}$), $\rho_\text{p}$ the density of the tracer particles ($\SI{1.04}{g.cm^{-3}}$), and $\mu_{\text{f}}$ the dynamic viscosity of the droplet liquid ($\sim 10^{-3} \SI{}{\pascal\second}$). The factor $(1+\rho_{\text{f}}/(2\rho_{\text{p}}))$ accounts for the added mass force \citep{Oliveira2015}.
The buoyancy-corrected Froude number is defined as $\mbox{\it Fr} \equiv \frac{u_{\text{max}}^2/r_{\text{c}}}{(1-\rho_{\text{f}}/\rho_{\text{p}})g}$, where $\rho_{\text{f}}$ is the density of the droplet liquid ($\SI{0.977}{g.cm^{-3}}$ for water and $\SI{0.789}{g.cm^{-3}}$ for ethanol) and $g$ is the gravitational acceleration.
The calculated values for $\mbox{\it St}$ and $\mbox{\it St}/\mbox{\it Fr}$ are $\sim 10^{-5}$ and $\sim 10^{-2}$ respectively, which indicate that the tracer particles are truly tracers for the liquid flow inside the evaporating droplets.

The side-view recording setup (section \ref{sec:exp:imageana}) was coupled to the micro-PIV setup to monitor the geometric shape evolution of the evaporating droplet synchronized with the flow inside. 

\subsection{Confocal Microscopy}
\label{sec:exp:confocal}
Unfortunately, the flow in ternary Ouzo droplets cannot be quantified using the micro-PIV technique, since the fluorescence signal of the tracer particles is concealed by the nucleation of oil microdroplets in its second phase \citep{Tan2016a}. The nucleated microdroplets themselves cannot be used as tracer particles because of their growing size and the non-uniform dispersion in the Ouzo droplet. However, a qualitative investigation of the flow in an Ouzo droplet can be performed by visualizing the phase separation and the movement of the oil microdroplets with confocal microscopy. To that end, real-time observations of Ouzo droplets were carried out by using an inverted Nikon A1 confocal laser scanning microscope system (Nikon Corporation, Tokyo, Japan) with a 20$\times$ magnification dry objective (CFI Plan Apochromat VC 20$\times$/0.75 DIC, numerical aperture $= 0.75$, working distance $= \SI{1.0}{\milli\meter}$). An Ouzo droplet consisting of trans-anethole, ethanol, and water was deposited on a hydrophobic glass coverslip with a thickness of $\SI{170}{\micro\meter}$. The initial volume of the droplets varied from $\SI{0.1}{\micro\liter}$ to $\SI{1.0}{\micro\liter}$. Trans-anethole oil was labeled by Nile Red (Sigma-Aldrich) which was excited by laser light with a wavelength of $\SI{561}{\nano\meter}$, while the water-ethanol mixture was labeled by Fluorescein 5(6)-isothiocyanate (FITC, Sigma-Aldrich) which was excited by laser light with a wavelength of $\SI{488}{\nano\meter}$. The scan started immediately after the droplet was deposited on the coverslip with a frame rate of $\SI{30}{\hertz}$ for scanning two-dimensional (2D) images under resonant mode. 

\section{Axisymmetric Investigation}
\label{sec:axisymm:axisymm}
In a first step in the model used in the numerical simulations, the drop is assumed to be axisymmetric. Although the flow velocity investigation later on in section \ref{sec:threedim:threedim} shows that this symmetry is broken, the assumption of axisymmetry drastically reduces the computation time. As shown by \citet{Diddens2017a}, the exact details of the flow in a droplet are not relevant for macroscopic quantities as e.g. the volume evolution $V(t)$, whenever the flow is driven by an intense Marangoni flow. This fact allows to discuss e.g. the influence of the latent heat of evaporation on the droplet evolution within an axisymmetric model.

\subsection{Axisymmetric Finite Element Method}
\label{sec:axisymm:modeldesc}
\begin{figure}\centering\includegraphics[width=1\textwidth]{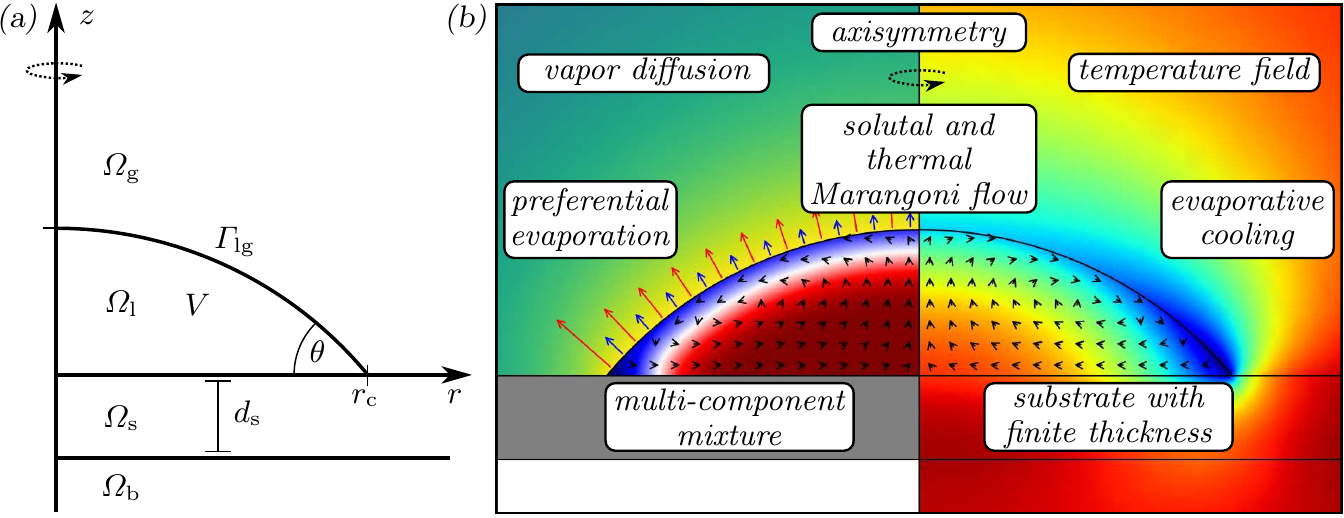}
\caption{(a) The problem is expressed by axisymmetric cylindrical coordinates. The individual domains $\Omega_{\text{g}}$, $\Omega_{\text{l}}$, $\Omega_{\text{s}}$ and $\Omega_{\text{b}}$ are the gas phase, the liquid droplet, the substrate with finite thickness $d_{\text{s}}$ and the air below the substrate, respectively. (b) The model solves the coupled processes of vapor-diffusion-limited mixture evaporation, multi-component flow with composition-dependent quantities and driven by solutal and thermal Marangoni flow, as well the temperature field. }
\label{fig:axisymm:scheme}
\end{figure}

In the following an outline of the axisymmetric finite element method is given. A detailed description of the model can be found in \citet{Diddens2017b}.
As depicted in \figurename~\ref{fig:axisymm:scheme}, the space, represented by axisymmetric cylindrical coordinates $(r,z)$, is separated into individual subdomains, namely the gas phase $\Omega_{\text{g}}$, the droplet liquid $\Omega_{\text{l}}$, the substrate $\Omega_{\text{s}}$ with a finite thickness $d_{\text{s}}$ and the air below the substrate $\Omega_{\text{b}}$. The droplet shape, defined by the liquid-gas interface $\Gamma_{\text{lg}}$, is assumed to be always in a spherical-cap shape. According to the ellipticity measurements of \citet{Tan2016a}, this is a good approximation, at least as long as the oil ring in case of the Ouzo droplet is not taken into account. 

The following derivation of the model considers the Ouzo droplet, but it can easily be simplified to the pure water droplet and the binary water-ethanol droplet. The composition-dependence of the physical fluid properties, i.e. viscosity $\mu$, diffusivity $D_{}$, mass density $\rho$, surface tension $\sigma$ and thermodynamic activities, are locally considered by fitting experimental data for water-ethanol mixtures \citep{Vazquez1995a,Gonzalez2007a,Parez2013a} and using thermodynamic models \citep{Zuend2008a,Zuend2011a}. Due to the small initial concentration of anise oil in the Ouzo mixture, its influence on the liquid properties is not considered. The temperature-dependence of the surface tension is also considered based on the experimental data of \citet{Vazquez1995a} to account for thermal Marangoni flow. Plots of all these relations can be found in \citet{Tan2016a} and \citet{Diddens2017a}.

In the gas phase, the diffusion equation for the vapor concentration $c_\alpha$ with $\alpha=\text{e}$ and $\alpha=\text{w}$ for ethanol and water, respectively, has to be solved. With the vapor diffusivity $D_{\alpha}^\text{g}$ of $\alpha$ in air, this reads
\begin{equation}
\partial_t c_\alpha = D_{\alpha}^\text{g}\nabla^2 c_\alpha  \,.
\label{eq:axisymm:gasdiffueq}
\end{equation}
The anise oil is assumed to be non-volatile due to the low vapor pressure of trans-anethole. As boundary condition of \eqref{eq:axisymm:gasdiffueq} for $(r,z)\to \infty$, the ambient vapor concentrations $c_{\alpha,\infty}$ have to be imposed. While there is no ethanol vapor present far away from the droplet, i.e. $c_{\text{e},\infty}=0$, the water vapor concentration can be expressed by the ideal gas law:
\begin{equation}
c_{\text{w},\infty}=\phi\frac{M_{\text{w}}p_{\text{w,sat}}(T_{\infty})}{R T_\infty}\,.
\label{eq:axisymm:cwaterinfty}
\end{equation}
Here, $\phi$ is the relative humidity, $R$ is the ideal gas constant, $T_{\infty}$ is the room temperature and $p_{\text{w,sat}}$ is the saturation pressure, which temperature-dependence is given by the Antoine equation. At the liquid-gas interface $\Gamma_{\text{lg}}$, the vapor-liquid equilibrium concentration $c_{\alpha,\text{VLE}}$ is imposed, which can be calculated via Raoult's law, i.e. by
\begin{equation}
c_{\alpha,\text{VLE}}=\gamma_{\alpha} x_{\alpha} \frac{M_{\alpha}p_{\alpha\text{,sat}}(T)}{R T} \,.
\label{eq:axisymm:cnuvle}
\end{equation}
The vapor-liquid equilibrium couples the vapor concentration at the droplet interface to the liquid mole fraction $x_{\alpha}$, where non-idealities are considered by the activity coefficient $\gamma_{\alpha}$. Since \eqref{eq:axisymm:cnuvle} has to be evaluated at the local temperature, the evaporation rates are also coupled to the temperature field $T$. From the solution of \eqref{eq:axisymm:gasdiffueq}, one can directly calculate the diffusive vapor flux at the interface:
\begin{equation}
\boldsymbol{J}_{\alpha}^\text{g}=-D_{\alpha}^\text{g}\bnabla c_\alpha\bcdot\boldsymbol{n}_{\text{lg}}  \,.  
\label{eq:axisymm:diffuevap}
\end{equation}
The mass transfer rates $j_{\text{w}}^\text{lg}$ and $j_{\text{e}}^\text{lg}$ can be determined from the diffusive vapor fluxes via the coupled mass transfer jump conditions
\begin{equation}
j_{\alpha}^\text{lg}-\boldsymbol{J}_{\alpha}^\text{g}\bcdot\boldsymbol{n}_{\text{lg}}-y_{\alpha}^\text{g}\left(j_{\text{w}}^\text{lg}+j_{\text{e}}^\text{lg}\right)=0\,,
\label{eq:axisymm:jumpcondgas}
\end{equation}
where $y_{\alpha}^\text{g}$ is the mass fraction of $\alpha$-vapor in the gas phase. By assuming constant gas density $\rho^\text{g}$, one can approximate $y_{\alpha}^\text{g}=c_{\alpha,\text{VLE}}/\rho^\text{g}$.

In the droplet, the spatio-temporal liquid composition is governed by the convection-diffusion equations for the liquid mass fractions $y_{\alpha}$, i.e. 
\begin{equation}
\rho\left(\partial_t y_{\alpha} + \boldsymbol{u}\bcdot\bnabla y_{\alpha} \right) = \bnabla\bcdot\left(\rho D_{} \bnabla y_{\alpha}\right)-\boldsymbol{J}_{\alpha}\bcdot\boldsymbol{n}_{\text{lg}}\delta_{\text{lg}}\,.
\label{eq:axisymm:strongcompo}
\end{equation}
For simplicity, the composition-dependence of the mass density $\rho$ is approximated by the spatially averaged composition. This assumption has been validated in \citet{Diddens2017b} and simplifies Eq. \eqref{eq:axisymm:strongcompo} as well as the momentum equation. At the liquid-gas interface, a source/sink term is imposed via the interface delta function $\delta_{\text{lg}}$. The corresponding flux $\boldsymbol{J}_{\alpha}$ is obtained by the counterpart of Eq. \eqref{eq:axisymm:jumpcondgas} for the liquid phase, i.e.
\begin{equation}
j_{\alpha}^\text{lg}-\boldsymbol{J}_{\alpha}\bcdot\boldsymbol{n}_{\text{lg}}-y_{\alpha}\left(j_{\text{w}}^\text{lg}+j_{\text{e}}^\text{lg}\right)=0\,.
\label{eq:axisymm:jumpcondliquid}
\end{equation}
The mass fraction of the non-volatile anise oil is not solved explicitly, but calculated via $y_{\text{a}}=1-y_{\text{e}}-y_{\text{w}}$.

Due to the latent heat of evaporation $\Lambda_{\alpha}$, the interface is cooled down by the evaporative mass fluxes $j_{\alpha}^\text{lg}$. This effect is taken into account by considering the convection-diffusion equation for the temperature $T$, i.e.
\begin{equation}
\rho{}c_p\left(\partial_t T + \boldsymbol{u}\bcdot\bnabla{}T\right)=\bnabla\bcdot(\lambda\bnabla T) - \delta_{\text{lg}} \left( j_{\text{w}}^\text{lg}\Lambda_{\text{w}}+ j_{\text{e}}^\text{lg}\Lambda_{\text{e}} \right)\,.
\label{eq:axisymm:tempfield}
\end{equation}
The mass density $\rho$, the specific heat capacity $c_p$ and the thermal conductivity $\lambda$ are different on the individual subdomains. In the droplet, the values furthermore depend on the local mixture composition according to the experimental data of \citet{Grolier1981a} and \citet{Yano1988a}. 

Finally, the velocity $\boldsymbol{u}$ has to be solved. For the moment, the flow in the gas phase is disregarded. The influence of convective mass and thermal transport in the gas phase is discussed later on in section \ref{sec:axisymm:gasflow}. Due to the micrometer-sized droplet, the Bond number $\mbox{\it Bo}$ and the capillary number $\mbox{\it Ca}$ are small, i.e. $\ll 1$ ($\mbox{\it Bo}<0.08$ and $\mbox{\it Ca}<3{\cdot}10^{-4}$ holds for all droplets discussed in this article). While the former allows to disregard gravitational effects, the latter ensures that the flow in the droplet is dictated by the surface tension. Thus, it is assumed that the droplet is always in an axisymmetric spherical cap shape. Thereby, it is possible to calculate the normal interface velocity $\boldsymbol{u}_{\text{lg}}\bcdot\boldsymbol{n}_{\text{lg}}$ along the entire interface directly from the evolution of the volume $V(t)$ and the contact angle $\theta(t)$. The volume loss $\dot{V}(t)$ can easily be calculated by integrating the evaporation rates over the liquid-gas interface and by taking the spatially averaged composition-dependence of the mass density into account. The contact angle $\theta$ can be imposed from fitted experimental data for $\theta(t)$ or $\theta(V)$ or, alternatively, obtained from an appropriate model for the composition-dependent contact angle evolution (cf. e.g. \citet{Diddens2017b}).

The normal interface velocity defines via the kinematic boundary condition the normal flow velocity $\boldsymbol{n}_{\text{lg}}\bcdot\boldsymbol{u}$ in the droplet, i.e. 
\begin{equation}
\boldsymbol{n}_{\text{lg}}\bcdot\boldsymbol{u}=\boldsymbol{n}_{\text{lg}}\bcdot\boldsymbol{u}_{\text{lg}}+\frac{1}{\rho}\left(j_{\text{w}}^\text{lg}+j_{\text{e}}^\text{lg}\right)=:u_n\,.
\label{eq:axisymm:normvelobc}
\end{equation}
With the stress tensor 
\begin{equation}
\mathsfbi{T}(\boldsymbol{u})=-p\mathsfbi{I} +\mu\left( \bnabla \boldsymbol{u}+(\bnabla \boldsymbol{u})^{\text{t}} \right)\,,
\label{eq:axisymm:sigmatens}
\end{equation}
the tangential velocity component is subject to the shear stress boundary condition
\begin{equation}
\boldsymbol{n}_{\text{lg}}\bcdot\mathsfbi{T}\bcdot\boldsymbol{t}_{\text{lg}} - \boldsymbol{n}_{\text{lg}}\bcdot\mathsfbi{T}^{\text{g}}\bcdot\boldsymbol{t}_{\text{lg}}=\boldsymbol{t}_{\text{lg}}\bcdot \bnabla_{\text{lg}}\sigma\,.
\label{eq:axisymm:marashear}
\end{equation}
Due to the small viscosity ratio between gas and liquid phase, i.e. $\mu^\text{g}\ll\mu$, and the continuous tangential velocity at the interface, the contribution of the shear stress in the gas phase, i.e. the term $\boldsymbol{n}_{\text{lg}}\bcdot\mathsfbi{T}^{\text{g}}\bcdot\boldsymbol{t}_{\text{lg}}$, can be neglected. This assumption is validated later on in section \ref{sec:axisymm:gasflow}.
Since the surface tension $\sigma$ is a function of the composition and the temperature, solutally and thermally driven Marangoni flow has to be expected.

Inside the droplet, a slightly modified Stokes flow is solved:
\begin{equation}
-\bnabla\bcdot\mathsfbi{T}(\boldsymbol{u})=0 \qquad\text{and}\qquad \bnabla\bcdot \boldsymbol{u}=-\frac{\partial_t\rho}{\rho}
\label{eq:axisymm:strongstokes}
\end{equation}
along with the no-slip boundary condition $\boldsymbol{u}=0$ at the substrate $z=0$.

For the implementation of the finite element method, the governing equations are converted to corresponding weak forms \citep{Diddens2017b} and solved with the help of the finite element package \textsc{FEniCS} \citep{bookLogg2012a}.

\subsection{Pure Water Droplet} 

\begin{figure}
\begin{minipage}{0.625\textwidth} 
\frame{\includegraphics[width=1\textwidth]{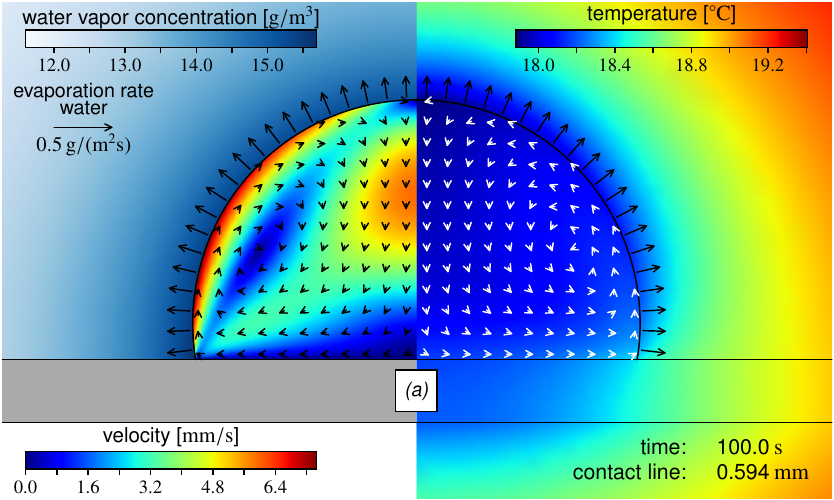}}
\frame{\includegraphics[width=1\textwidth]{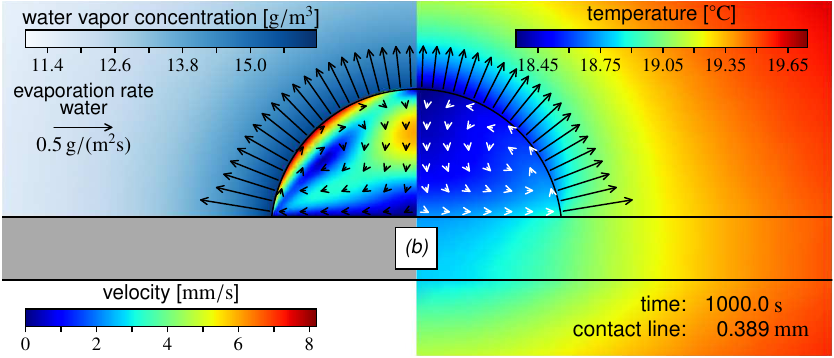}}
\end{minipage}\hfill
\begin{minipage}{0.35\textwidth} 

\includegraphics[width=1\textwidth]{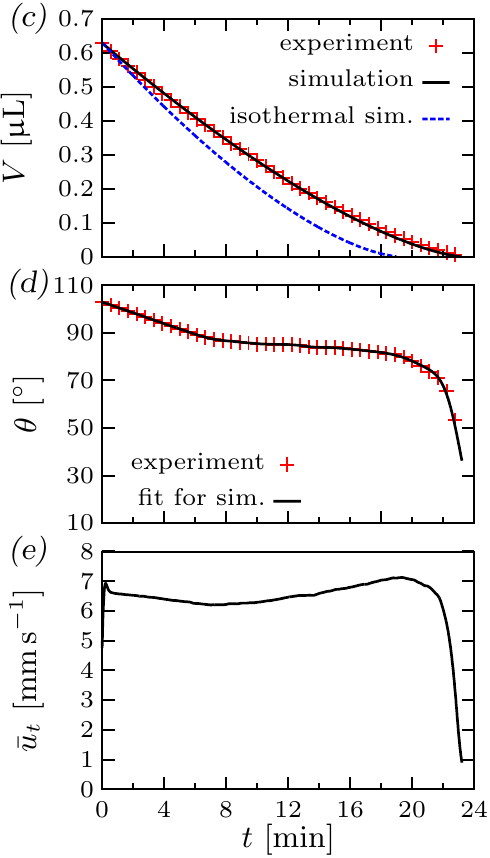}
\end{minipage}
  \caption{(a-b) Snapshots of a pure water droplet with velocity field in the droplet and vapor concentration in the gas phase (left) and temperature field (right) at two different times $t=\SI{100}{\second}$ (a) and $t=\SI{1000}{\second}$ (b). A corresponding movie is available as supplementary material. (c) Volume evolution of the experiment and according to simulations with and without considering thermal effects. (d) Contact angle of the experiment and corresponding fit, which is used as input parameter for the simulation. (e) Averaged tangential fluid velocity at the interface. } \label{fig:axisymm:purewater}  \end{figure} 

To validate the model on the basis of the simplest case, the experimental setup was used to measure the evolution of a $\SI{0.63}{\micro\liter}$ pure water droplet at ambient temperature $T_{\infty}=\SI{20.4}{\celsius}$ and relative humidity $\phi=\SI{54}{\percent}$. The case of pure water can easily be simulated by setting $y_{\text{w}}=1$ in the model equations. The experimental data is depicted along with the corresponding simulation data in \figurename~\ref{fig:axisymm:purewater}. The right side of the snapshots (a) and (b) show the temperature field of the droplet and its surrounding. Due to the thin substrate, the latent heat of evaporation leads to an intense cooling, which can be more than $\SI{3}{\kelvin}$. From the volume evolution $V(t)$ depicted in (c), it is apparent how the cooling influences the evaporation rate: If the latent heat and the corresponding reduction of the vapor pressure are considered, the numerical simulation perfectly agrees with the experimental data. For comparison, also the corresponding isothermal simulation, i.e. without considering thermal effects, is indicated. This simulation, which resembles the result of the model equation by \citet{Popov2005a}, predicts a noticeably faster drying than the experiment. Thus, even for pure droplets evaporating at room temperature on a glass substrate, thermal effects cannot be disregarded, provided the substrate is thin, i.e. comparable with the droplet size, with air below.

The contact angle evolution in \figurename~\ref{fig:axisymm:purewater}(d) shows the typical stick-slip behavior, i.e. an initial decrease of $\theta(t)$ at constant contact line radius $r_\text{c}$, followed by a constant contact angle with a receding contact line $r_\text{c}(t)$. While this behavior could be incorporated into the simulation by introducing a receding contact angle, the contact line dynamics of mixture droplets are more rich and complicated to account for with an accurate model. To that end, we have simply fitted the contact angle measured experimentally and used it as an input parameter throughout the simulation.

As can be seen from the left side of the snapshots (a) and (b), the simulation predicts a strong thermal Marangoni flow from the contact line along the liquid-gas interface towards the apex. The corresponding temporal evolution can be inferred from \figurename~\ref{fig:axisymm:purewater}(e), where the average tangential fluid velocity at the interface
\begin{equation}
\bar{u}_{t}(t)=\frac{1}{ \int_{\Gamma_{\text{lg}}} r{\rm d}l} \int_{\Gamma_{\text{lg}}}  \boldsymbol{u}\bcdot \boldsymbol{t}_{\text{lg}} \: r{\rm d}l
\label{eq:axisymm:meantangvelo}
\end{equation}
is plotted. The tangent $\boldsymbol{t}_{\text{lg}}$ is defined to point along the interface towards the apex. One can infer that the simulation predicts a persistent thermal Marangoni flow throughout the entire lifetime of the droplet. In experiments, the observed thermal Marangoni flow in water droplets is usually much slower than the theoretical predictions \citep{Hu2005b}. We discuss this issue later on in section \ref{sec:threedim:micropiv}.

\subsection{Binary Water-Ethanol Droplet}

\begin{figure}
\begin{minipage}{0.635\textwidth} 
\frame{\includegraphics[width=1\textwidth]{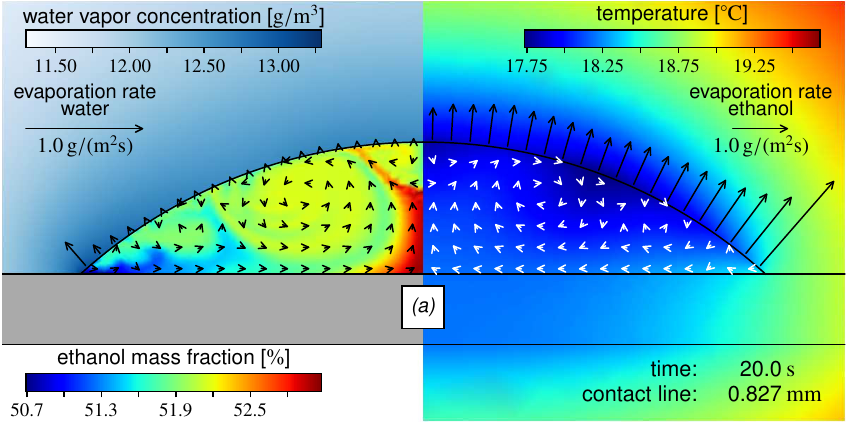}}
\frame{\includegraphics[width=1\textwidth]{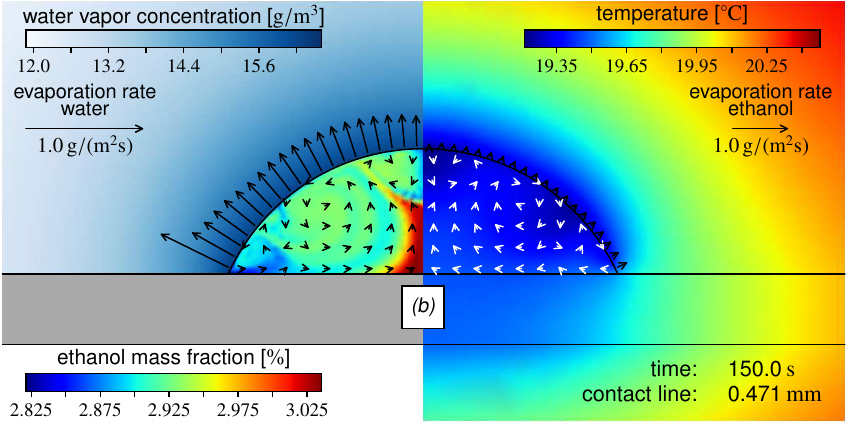}}
\frame{\includegraphics[width=1\textwidth]{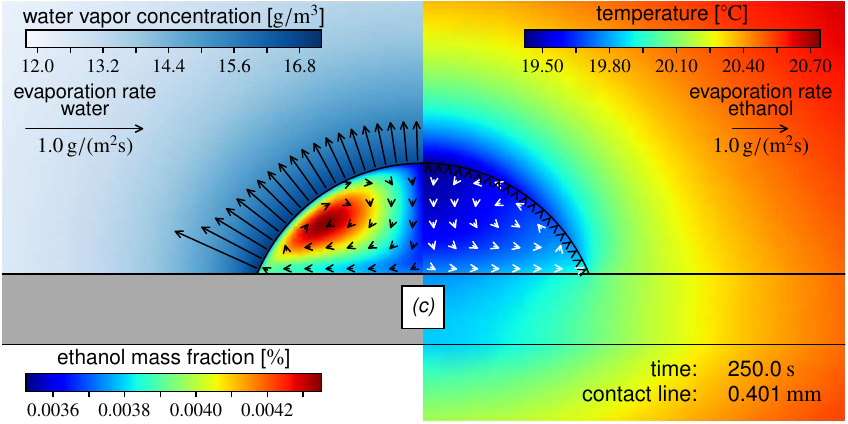}}
\end{minipage}\hfill
\begin{minipage}{0.34\textwidth} 

\includegraphics[width=1\textwidth]{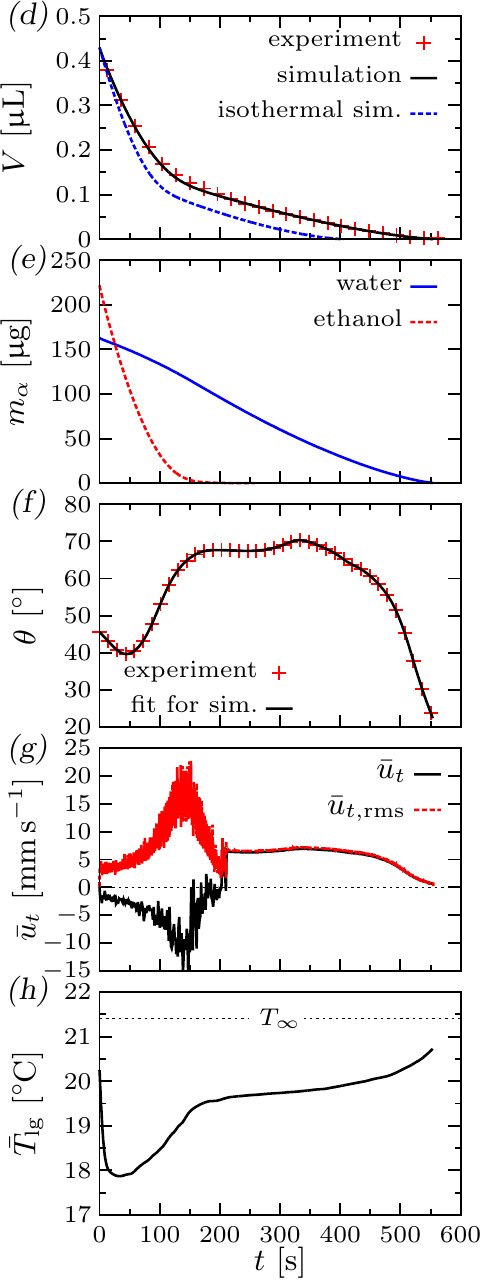}
\end{minipage}
  \caption{(a-c) Snapshots of a binary water-ethanol droplet with ethanol mass fraction $y_{\text{e}}$ in the liquid phase and water vapor concentration in the gas phase (left) and temperature field (right) at three different times $t=\SI{20}{\second}$ (a), $t=\SI{150}{\second}$ (b) and $t=\SI{250}{\second}$ (c). The arrows at the interface indicate the evaporation rates of water (left) and ethanol (right). A corresponding movie is available as supplementary material. (d) Volume evolution according to experiment, full simulation and isothermal simulation. (e) Partial masses of both components in the drop. (f) Experimentally obtained contact angle and corresponding fit for the model. (g) Mean tangential fluid velocity at the interface and its rms-average. (h) Average interface temperature. } \label{fig:axisymm:binary}  \end{figure} 

As a next step, the model is validated by a binary water-ethanol droplet at ambient temperature $T_{\infty}=\SI{21.4}{\celsius}$ and relative humidity $\phi=\SI{55}{\percent}$. Due to practical reasons in the experiment, it took some seconds between the deposition of the droplet and the first snapshot at $t=\SI{0}{\second}$. Since the average composition changes within this offset time, the initial composition in the simulation has a slightly lower ethanol content (\SI{57.7}{\wtpercent}) than specified in section \ref{sec:exp:composition}. This particular correction was determined by extrapolating the numerically obtained initial compositional rate of change over the offset time.

Representative snapshots of the simulation are shown in \figurename~\ref{fig:axisymm:binary}(a-c). Initially, the droplet is rather flat and both components evaporate with the typical singularity near the contact line. In combination with the fact that water (evaporation rate by arrows on the left side) has a lower volatility than ethanol (arrows on the right side), this leads to an enhanced water concentration near the contact line. The resulting surface tension gradient drives an intense solutal Marangoni flow, which breaks up into chaos due the Marangoni instability of water-ethanol mixtures \citep{Machrafi2010a}.

In the initial regime, thermal Marangoni flow is almost irrelevant, but the incorporation of the latent heat of evaporation is still important, as it affects again the volume evolution, which is depicted in \figurename~\ref{fig:axisymm:binary}(d). As for the case of pure water, the volume $V(t)$ of the simulation agrees perfectly with the experimental data, but only if the evaporative cooling is considered. In the case of a binary mixture, the evaporative cooling has another effect, namely the suppression of the water evaporation at initial stages: Due to the highly volatile ethanol, the interface is cooled down to a temperature, at which the vapor-liquid equilibrium for the water vapor concentration at the interface almost reaches the level of the vapor concentration in the ambient. At a higher relative humidity, this coupling between the evaporation rates and the temperature would even induce an initial condensation of water \citep{Diddens2017b}.

The two distinct slopes in the volume curve $V(t)$, which we already observed and explained in \citet{Tan2016a}, result from the preferential evaporation of ethanol in the initial regime followed by the slower evaporation of the remaining water residual at later times. This interpretation is validated by resolving the evolution into the partial masses of water and ethanol as shown in \figurename~\ref{fig:axisymm:binary}(e). It is apparent that the evaporation rate of water, i.e. the slope of the partial mass $\dot{m}_\text{w}(t)$, is slightly reduced as long as ethanol is present. This is due to Raoult's law \eqref{eq:axisymm:cnuvle}, but is also caused by the reduction of the vapor pressure due to the additional cooling stemming from the evaporation of ethanol.

The contact angle was again fitted from experimental data and used as input parameter for the simulation. The data for $\theta(t)$ depicted in \figurename~\ref{fig:axisymm:binary}(f) reveals an initial decrease due to a pinned contact line followed by an intense increase, which can be understood by considering the composition-dependence of the surface tension in the Young-Laplace equation.

At later times, when virtually all ethanol has evaporated, a flow transition occurs (cf. \figurename~\ref{fig:axisymm:binary}(c)): The chaotic solutal Marangoni flow suddenly switches over to regular thermal Marangoni flow as for the pure water droplet of \figurename~\ref{fig:axisymm:purewater}. The abrupt transition can also be inferred from the plot of the average tangential velocity $\bar{u}_{t}(t)$ in \figurename~\ref{fig:axisymm:binary}(g). While the chaotic solutal flow induces a net flow towards the contact line, $\bar{u}_{t}<0$, it suddenly converges to the regular thermal Marangoni flow towards the apex, i.e. $\bar{u}_{t}>0$. Additionally, to estimate the average flow speed at the interface without respecting the local direction, the rms-averaged tangential fluid velocity
\begin{equation}
\bar{u}_{t,\text{rms}}(t)=\sqrt{\frac{1}{ \int_{\Gamma_{\text{lg}}} r{\rm d}l} \int_{\Gamma_{\text{lg}}}  \left(\boldsymbol{u}\bcdot \boldsymbol{t}_{\text{lg}}\right)^2 \: r{\rm d}l }
\label{eq:axisymm:meantangvelorms}
\end{equation}
is plotted in \figurename~\ref{fig:axisymm:binary}(g). It is apparent that the flow is most intense in the limit of rather dilute ethanol. In this region, the compositional surface tension gradient, i.e. $\partial \sigma/\partial{}y_{\text{e}}$, has its maximum and hence drives the most intense solutal Marangoni flow.

Finally, in \figurename~\ref{fig:axisymm:binary}(h), the average temperature of the liquid-gas interface $\bar{T}_{\text{lg}}$ is depicted. Again, cooling can be up to $\SI{3.5}{\kelvin}$ and is most pronounced during the initial regime, where both components evaporate. Although water has a $2.7$ times higher specific latent heat than ethanol, most of the initial cooling is induced by the up to $9$ times faster evaporation rate of ethanol.

\subsection{Influence of Convective Transport in the Gas Phase}
\label{sec:axisymm:gasflow}
Before addressing the ternary Ouzo droplet, the flow in the gas phase is investigated. Until now, this flow has been disregarded, but since it generates convective vapor and thermal transport, it can influence the evaporation rates and the temperature field. 
The flow in the gas phase can most generally be driven by four mechanisms, namely forced convection, natural convection, Marangoni convection and Stefan flow. While forced convection can be ruled out due to the geometry of the experimental setup, the mass density gradient required for natural convection can be caused by a thermal gradient as well as by vapor concentration gradients. Since the temperature at the droplet is lower than the ambient temperature, thermally driven natural convection can be ruled out. 
To estimate the influence of solutally induced natural convection, we define the solutal Rayleigh number
\begin{equation}
\mbox{\it Ra}_\alpha=\frac{g\beta_{\alpha}\left(c_{\alpha,\text{VLE}} - c_{\alpha,\infty} \right)r_\text{c}^3}{\nu D_{\alpha}^\text{g}} \,,
\label{eq:axisymm:rayleighsol}
\end{equation}
where $g$ is the acceleration due to gravity, $\nu$ the kinematic viscosity of the gas phase (assumed to be independent of the vapor concentration) and $\beta_{\alpha}=(\partial \rho^\text{g}_{}/\partial{}c_{\alpha})/\rho^\text{g}_{}$ is the solutal expansion coefficient, which can be calculated by the ideal gas law.
The solutal Rayleigh numbers with the largest modulus of the simulations discussed so far read $\mbox{\it Ra}_{\text{w}}\sim 1$ and $\mbox{\it Ra}_{\text{e}}\sim -10$ for water vapor and ethanol vapor, respectively. Note that the different signs originate from the fact that water vapor is less dense than dry air, whereas ethanol vapor is more dense. However, according to \citet{Dietrich2016a}, an influence of solutally driven natural convection on the evaporation rate can be ruled out at these Rayleigh numbers.

Since the tangential velocity is continuous at the interface, the fast Marangoni flow will be also present in the gas phase. Furthermore, the density difference of liquid and vapor leads to a discontinuous jump in the normal velocity component, which constitutes the so-called Stefan flow. To investigate the influence of these effects, the Stokes flow is solved in the gas phase as well, where the normal component of the gas velocity is imposed analogously to \eqref{eq:axisymm:normvelobc} and the tangential component is imposed to be continuous with its counterpart in the liquid. Furthermore, the convection-diffusion equations \eqref{eq:axisymm:gasdiffueq} and \eqref{eq:axisymm:tempfield} are augmented with the corresponding convective term in the gas phase and the additional convective vapor transport at the interface, i.e. $c_{\alpha,\text{VLE}}(\boldsymbol{u}_{\text{g}}-\boldsymbol{u}_{\text{lg}})\bcdot\boldsymbol{n}_{\text{lg}}$, is added to the evaporative mass fluxes.

\begin{figure}
\begin{minipage}{0.62\textwidth} 
\frame{\includegraphics[width=1\textwidth]{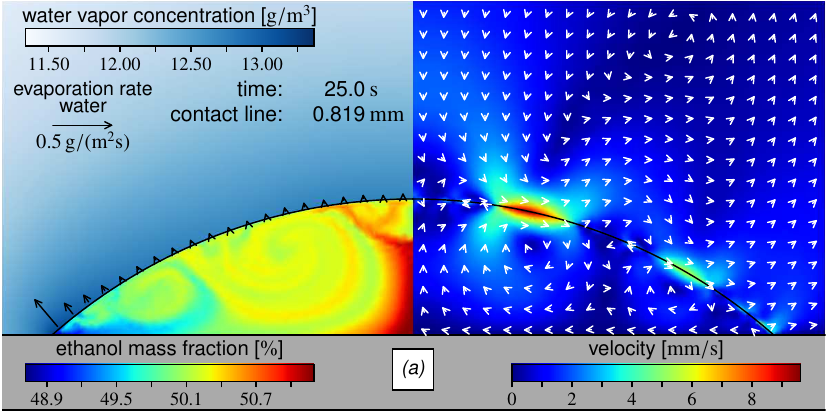}}
\frame{\includegraphics[width=1\textwidth]{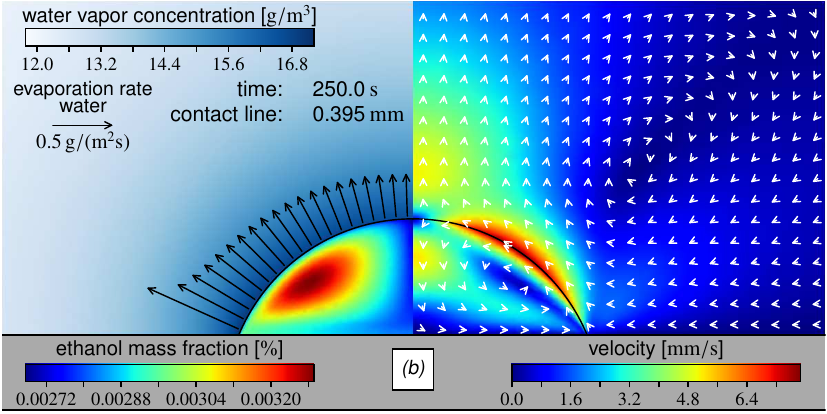}}
\end{minipage}\hfill
\begin{minipage}{0.35\textwidth} 

\includegraphics[width=1\textwidth]{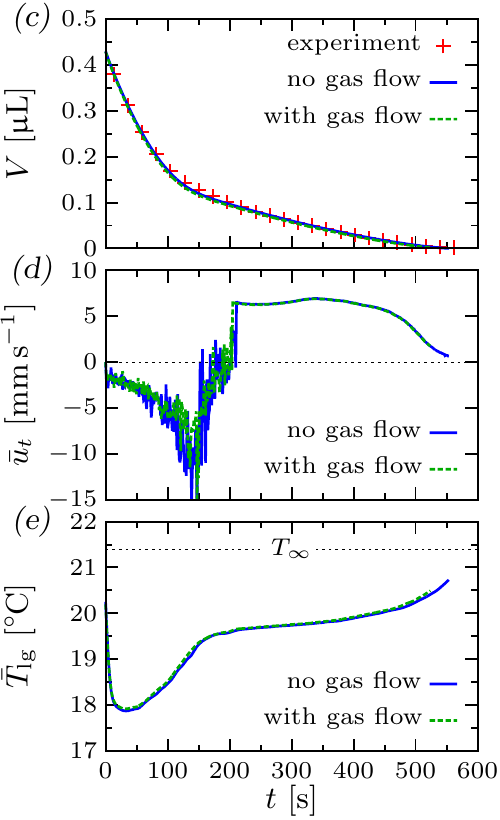}
\end{minipage}
  \caption{(a-b) Same as in \figurename~\ref{fig:axisymm:binary} at two different times $t=\SI{25}{\second}$ (a) and $t=\SI{250}{\second}$ (b), but now considering in the calculation and showing in the picture the flow in the gas phase (depicted on the right side). (c-e) Comparison of relevant quantities obtained by simulations with and without flow in the gas phase, revealing hardly any difference. } \label{fig:axisymm:airflow}  \end{figure} 

In \figurename~\ref{fig:axisymm:airflow}(a,b), the simulation of the binary water-ethanol droplet of \figurename~\ref{fig:axisymm:binary} with consideration of Marangoni and Stefan flow in the gas phase is depicted. From the velocity field on the right side of the snapshots, it is apparent how the Marangoni flow creates vortices in the gas phase, which are initially chaotic and become regular once the ethanol has evaporated. The contribution of the Stefan flow, i.e. the normal velocity jump, is barely visible, since the typical Stefan flow velocity is only about \SI{1}{\milli\meter\per\second}. As can be inferred from \figurename~\ref{fig:axisymm:airflow}(c-e), the additional convective transport of vapor and energy has virtually no influence on the volume evolution, the fluid velocity and the interfacial temperature. Hence, disregarding the flow in the gas phase is justified for the discussed cases.

Furthermore, the simulations with flow in the gas phase allow to validate the assumption of omitting the shear stress term $\boldsymbol{n}_{\text{lg}}\bcdot\mathsfbi{T}^{\text{g}}\bcdot\boldsymbol{t}_{\text{lg}}$ in \eqref{eq:axisymm:marashear}: The ratio of the shear stresses in the gas phase and in the liquid phase is indeed below $\SI{2}{\percent}$.

\subsection{Ternary Ouzo Droplet}
\begin{figure}
\begin{minipage}{0.635\textwidth} 
\frame{\includegraphics[width=1\textwidth]{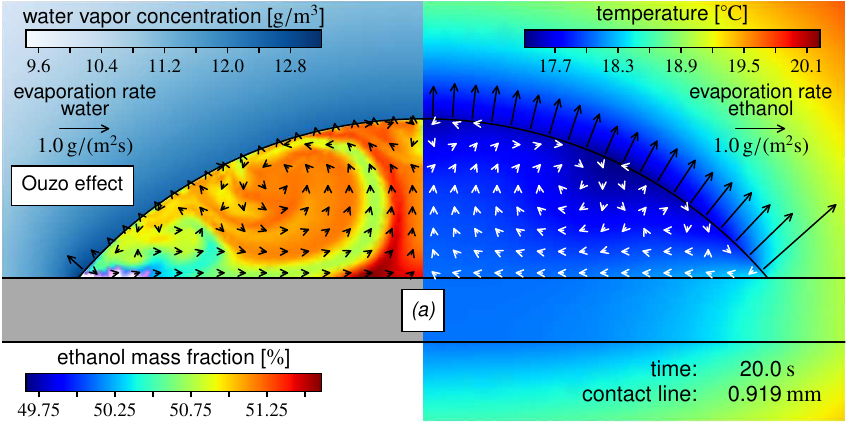}}
\frame{\includegraphics[width=1\textwidth]{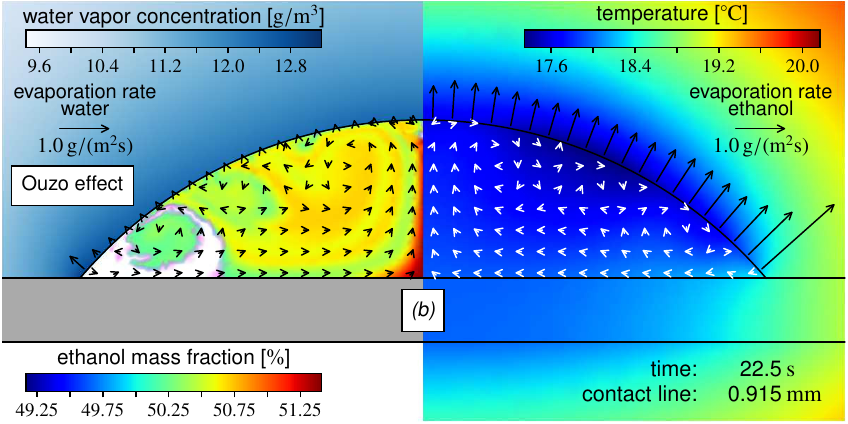}}
\frame{\includegraphics[width=1\textwidth]{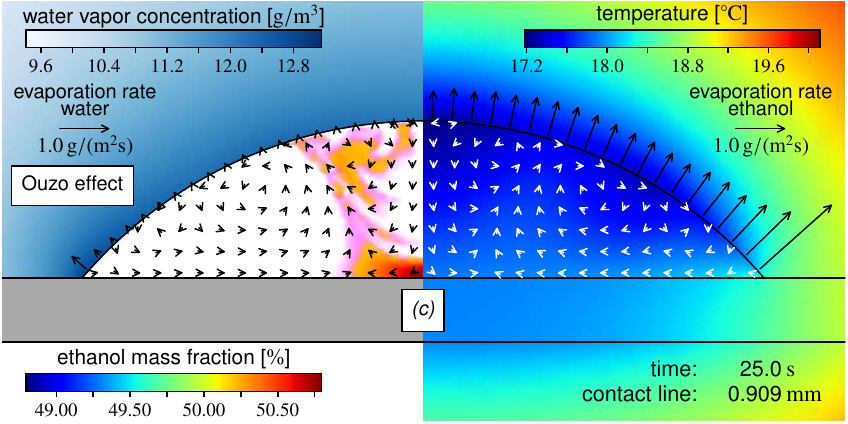}}
\end{minipage}\hfill
\begin{minipage}{0.34\textwidth} 

\includegraphics[width=1\textwidth]{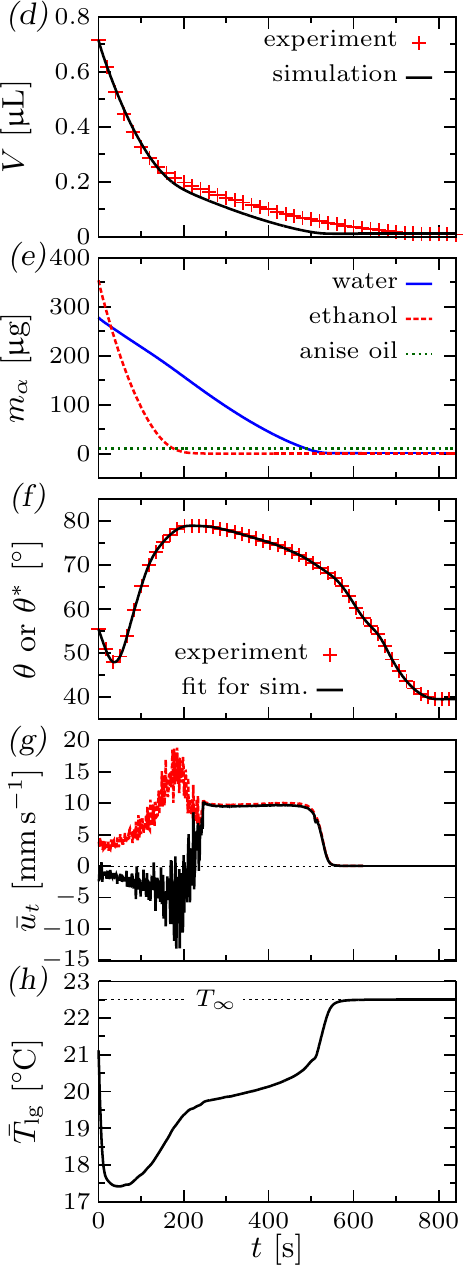}
\end{minipage}
  \caption{(a-c) Snapshots of a ternary Ouzo drop with ethanol mass fraction $y_{\text{e}}$ in the liquid phase and water vapor concentration in the gas phase (left) and temperature field (right) at three different times $t=\SI{20}{\second}$ (a), $t=\SI{22.5}{\second}$ (b) and $t=\SI{25}{\second}$ (c). Regions where the Ouzo effect occurs are shaded white on the left side. The arrows at the interface indicate the evaporation rates of water (left) and ethanol (right). A corresponding movie is available as supplementary material. (d) Volume evolution according to experiment and the full simulation. (e) Partial masses of all components in the drop. (f) During the evolution, the droplet shape temporarily deviates from a spherical-cap shape due to the presence of the oil ring. During this period, the contact angle in the model was set to the contact angle $\theta^*$ of the spherical cap-shaped upper part. Otherwise, the usual contact angle $\theta$ was imposed. (g) Mean tangential fluid velocity at the interface and its rms-average. (h) Average interface temperature. } \label{fig:axisymm:ouzo}  \end{figure} 

Now that the model has been validated based on a pure water and a binary water-ethanol droplet, we focus on a ternary Ouzo droplet at $T_{\infty}=\SI{22.5}{\celsius}$ and $\phi=\SI{40}{\percent}$ in the following. To that end, we took the experimental data of our previous article \citep{Tan2016a} and performed corresponding simulations with the finite element method described in section \ref{sec:axisymm:modeldesc}. Different from the lubrication theory model used in our previous article \citep{Tan2016a}, the present model takes thermal effects into account and is not subject to the limitation of the lubrication approximation where the contact angle has to be small. In the previous article, we furthermore had to adjust the relative humidity to reproduce the experimental data with the isothermal lubrication theory model in \citet{Tan2016a}. 

In \figurename~\ref{fig:axisymm:ouzo}(a-c), the numerically determined onset of the Ouzo effect is depicted (white regions on the left side). As in the experiment, the nucleation of oil microdroplets starts at about $t=\SI{20}{\second}$ at the contact line. Different from the predictions of the isothermal lubrication theory model \citep{Tan2016a}, the flow in the droplet is chaotic, which leads to a chaotic propagation of the phase separation front through the droplet. At $t\approx\SI{30}{\second}$, nucleation can occur everywhere in the droplet. 

As it is apparent from the volume evolution curve in \figurename~\ref{fig:axisymm:ouzo}(d), the initial slope matches the experimental results perfectly, but shows deviations at later times. Since the volume evolution of the pure water droplet and the binary water-ethanol droplet perfectly matches the experiments, the disagreement in the ternary case can definitely be attributed to the presence of the third component, namely the anise oil. While it is initially miscible, which is an assumption of the model, the simulation shows perfect agreement in the first regime. At later times, however, the oil ring starts to form, which changes the geometry from the typical spherical cap shape to a more complicated shape, i.e. a water droplet sitting on an oil ring. This leads to a reduction of the water-air interface area and thereby reduces the evaporation rate. The model cannot account for this change in geometry since it always assumes a perfect spherical cap shape and hence predicts a faster evaporation rate. Another possible factor that could contribute to the disagreement is the fact that the anise oil used is not a pure grade of trans-anethole and consists of about $\SI{10}{\percent}$ of unknown components \citep{Rodrigues2003a}. These could also influence the evaporation rate, e.g. by forming a shielding monolayer suppressing evaporation \citep{Langmuir1943a}.

The other quantities depicted in \figurename~\ref{fig:axisymm:ouzo}(e-h), e.g. the fluid velocity and temperature, are qualitatively similar to the case of the binary water-ethanol droplet of \figurename~\ref{fig:axisymm:binary}. While there are temporarily two contact angles in the experiment, i.e. the contact angle $\theta$ between the oil ring and the substrate and the contact angle $\theta^*$ of the spherical cap-shaped upper part of the droplet, the present simulation can only take a single contact angle. Here, $\theta^*$ was imposed as the contact angle in the simulation during the time the oil ring was present in the experiment.  

\section{Axial Symmetry Breaking}
\label{sec:threedim:threedim}
While the volume evolution predicted by the model has already been compared to the experimental data, a quantitative comparison of the flow velocity remains to be made.
However, for the investigation of the initially chaotic flow in the binary water-ethanol and in the ternary Ouzo droplet, an axisymmetric model clearly falls short, since the Marangoni instability will also induce compositional perturbations around the circumference of the droplet and drive Marangoni flow in this direction. This breaking of the axial symmetry for binary water-ethanol droplets has already been observed by \citet{Christy2011a}, \citet{Bennacer2014a} and \citet{Zhong2016a}; however, to our knowledge, it has not yet been studied by numerical simulations. In the following, we investigate this aspect of multi-component droplets by comparing experimental micro-PIV measurements with the results from the three-dimensional finite element method. Here, we focus on the binary water-ethanol droplet as depicted in \figurename~\ref{fig:axisymm:binary} for two reasons: On the one hand, the volume evolution shows perfect agreement for this droplet and, on the other hand, the presence of oil microdroplets in the Ouzo droplet hinders the use of the micro-PIV technique as described in section \ref{sec:exp:micropiv}. However, the flow in the Ouzo droplet can be qualitatively analyzed by confocal microscopy, which is done in section \ref{sec:confocal:confocal}.

\subsection{Three-Dimensional Model}
\label{sec:threedim:threedimfem}

\begin{figure}
\begin{minipage}{0.5\textwidth} 
\includegraphics[width=0.99\textwidth]{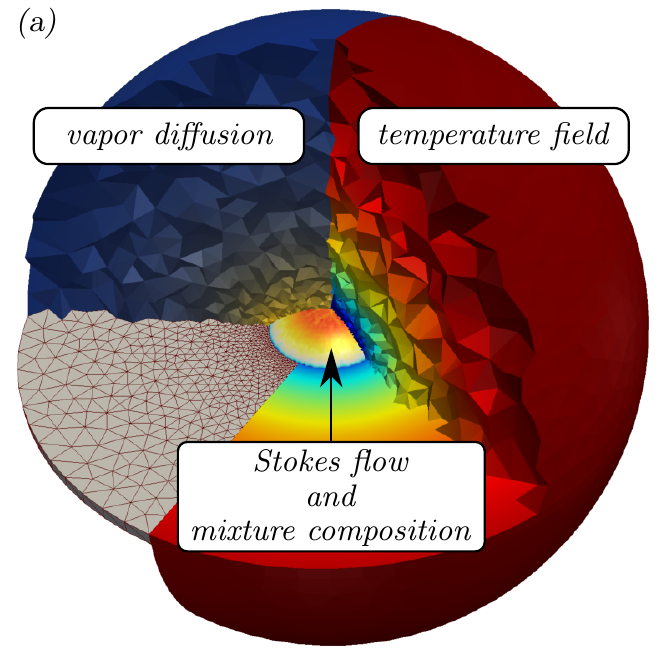}
\end{minipage}
\begin{minipage}{0.5\textwidth} 
\includegraphics[width=0.9\textwidth]{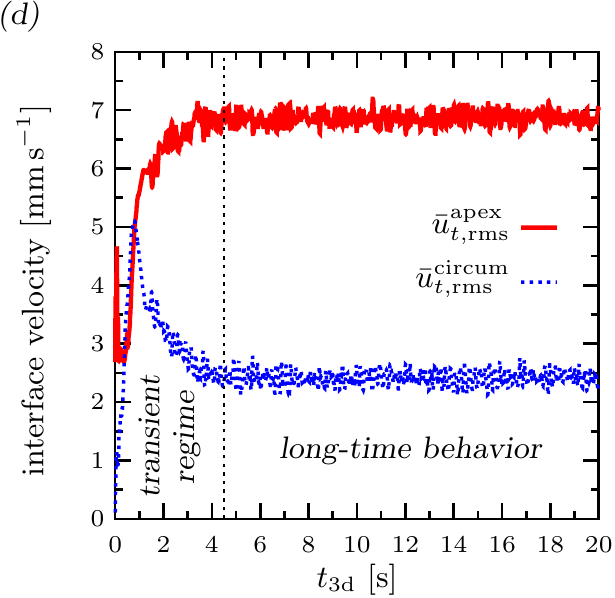}
\ \\ \ \\
\end{minipage}\\
\begin{minipage}{1.0\textwidth} 

\includegraphics[width=1\textwidth]{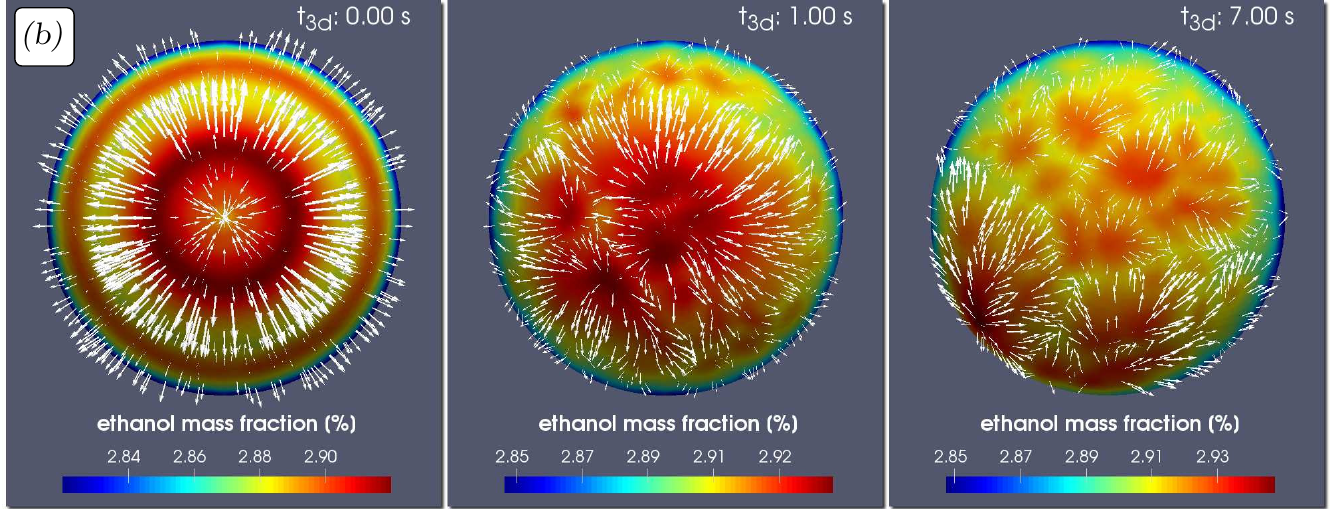}
\end{minipage}\\
\begin{minipage}{1.0\textwidth} 

\includegraphics[width=1\textwidth]{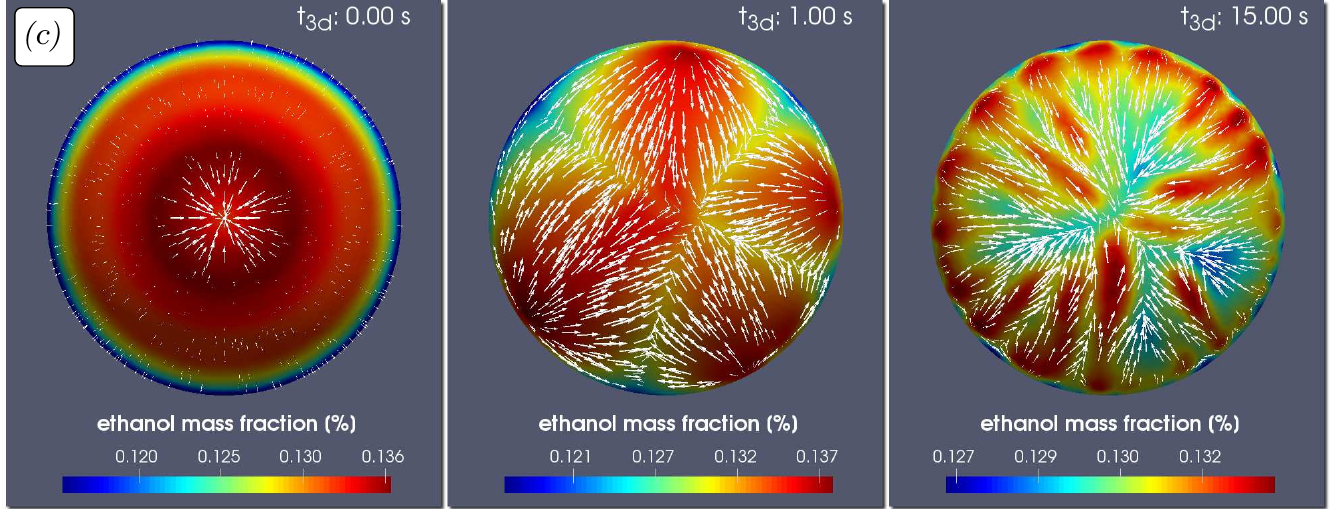}
\end{minipage}
  \caption{(a) View into a mesh for the three-dimensional model. (b) Axial symmetry breaking for snapshot time $t=\SI{150}{\second}$ at different three-dimensional simulation times $t_{\text{3d}}$. The droplet is depicted in top view showing the ethanol mass fraction at the interface. The composition and the tangential fluid velocity at the interface (white arrows) quickly break up into a chaotic, highly non-axisymmetric behavior with a cellular structure. (c) Same as (b), but for the snapshot time $t=\SI{200}{\second}$. Due to the low ethanol concentration, the break-up is slower. At later times, a cellular pattern can be observed along the rim, which is pulled in filaments towards the apex  by thermal Marangoni flow.  (d) Evolution of the rms-averaged tangential fluid velocity components $\bar{u}_{t,\text{rms}}^{\text{apex}}$ and $\bar{u}_{t,\text{rms}}^{\text{circum}}$ for the simulation of (c). After a transient regime, the dynamics end up in a chaotic long-time regime with converged mean values and standard deviations. } \label{fig:threedim:scheme}  \end{figure} 

To investigate the axial symmetry breaking within the framework of the model, it has to be generalized to three dimensions. Mathematically, this generalization from the two-dimensional axisymmetric model to a full three-dimensional variant is an easy step. One just has to transfer the weak forms \citep[cf.][]{Diddens2017b} to their corresponding three-dimensional equivalents. However, from the perspective of implementation, one faces the challenge of having to treat a free boundary problem. In the axisymmetric model, the movement of the sharp liquid-gas interface was made possible by shifting interfacial mesh nodes and performing a local reconstruction of the mesh whenever necessary to prevent the elements from collapsing. For the interpolation from the previous mesh to the reconstructed mesh, the supermesh method introduced by \citet{Farrell2009a} was utilized. Both the local mesh reconstruction and the supermesh method constitute a fundamental challenge to be generalized to three dimensions. To circumvent these problems, we focus on a static three-dimensional mesh instead. This approach does not allow to perform a single simulation over the entire lifetime of the droplet, but can still capture the three-dimensional flow statistics for a specific droplet composition. The use of a static mesh is possible, since the typical flow velocity in the droplet is several orders of magnitude larger than the movement of the interface. 

The procedure is as follows: For selected individual times $t$ of the axisymmetric simulation depicted in \figurename~\ref{fig:axisymm:binary}, we project the axisymmetric data on a corresponding three-dimensional mesh (cf. \figurename~\ref{fig:threedim:scheme}(a)). This is used as initial condition for the generalized three-dimensional model, which is integrated over another time $t_{\text{3d}}$ to obtain the characteristic three-dimensional flow at time instant $t$.  Again, the vapor diffusion equations for the volatile components are solved in the gas phase, the convection-diffusion equations for the composition and the Stokes flow are solved in the droplet, and the temperature field is solved in all domains, including the substrate and the air below it.
Since the mesh is static, the kinetic boundary condition has to be modified to $\boldsymbol{n}_{\text{lg}}\bcdot\boldsymbol{u}=0$. Furthermore, to capture the characteristic three-dimensional flow at time instant $t$, one has to ensure that the average composition is not changing within the simulation time $t_{\text{3d}}$ of the three-dimensional model. This can be achieved by augmenting the convection-diffusion equation \eqref{eq:axisymm:strongcompo} by a spatially uniform correction term which compensates for the evaporation, i.e.
\begin{equation}
\rho\left(\partial_{t_\text{3d}} y_{\alpha} + \boldsymbol{u}\bcdot\bnabla y_{\alpha} \right) = \bnabla\bcdot\left(\rho D_{} \bnabla y_{\alpha}\right)-\boldsymbol{J}_{\alpha}\bcdot\boldsymbol{n}_{\text{lg}}\delta_{\text{lg}} + \frac{1}{V}\int_{\Gamma_\text{lg}} \boldsymbol{J}_{\alpha}\bcdot{\rm d}\boldsymbol{A}\,.
\label{eq:threedim:strongcompo}
\end{equation}

The axisymmetry of the initial condition from the axisymmetric model can break up by the Marangoni instability. Azimuthal perturbations of the composition can arise, yielding to non-axisymmetric solutal Marangoni flow which in turn feeds back to the coupled dynamics of multi-component flow, evaporation and thermal effects. A representative example is shown in \figurename~\ref{fig:threedim:scheme}(b) for the snapshot time $t=\SI{150}{\second}$, which corresponds to the axisymmetric initial condition of \figurename~\ref{fig:axisymm:binary}(b). Within a very short three-dimensional simulation time $t_{\text{3d}}$, the axial symmetry of mixture composition and Marangoni flow is broken and the system exhibits highly non-axisymmetric dynamics with a cellular structure. 

When the ethanol concentration is lower and thermal Marangoni flow becomes relevant, the symmetry break-up is slower. This can be seen from the simulation at snapshot time $t=\SI{200}{\second}$ depicted in \figurename~\ref{fig:threedim:scheme}(c). In an initial transient regime, perturbations arise at the apex and near the contact line, which coalesce to four large cells. With ongoing simulation time, however, new perturbations arise at the contact line, which form small spots with enhanced ethanol concentration. Due to thermal Marangoni flow, filaments of enhanced ethanol concentration are pulled out of these spots towards the apex.
The spots near the contact line have a remarkable similarity to the hydrothermal waves in ethanol droplets on heated substrates as reported by \citet{Sobac2012a}. The fundamental difference is, however, that the hydrothermal waves are a result of an intense thermal Marangoni effect alone, whereas here a combination of thermal and solutal Marangoni flow induces these structures. 

For even lower ethanol concentrations, i.e. for snapshot times $t>\SI{225}{\second}$, the initial axisymmetry persists, i.e. a breaking of the axial symmetry cannot be observed.

To determine the typical three-dimensional flow characteristics, the simulation time $t_{\text{3d}}$ has to be sufficiently long. In particular, we are interested in the long-time regime only, i.e. not in the transient regime as e.g. depicted in the central picture of \figurename~\ref{fig:threedim:scheme}(c). To determine whether the simulation has already reached its long-time behavior, the rms-averaged tangential fluid velocity at the interface is investigated. In order to distinguish between axisymmetric behavior and broken symmetry later on, the rms is split into its two components, namely
\begin{align}
\bar{u}_{t,\text{rms}}^{\text{apex}}(t)&=\sqrt{\frac{1}{ \int_{\Gamma_{\text{lg}}} {\rm d}A} \int_{\Gamma_{\text{lg}}}  \left(\boldsymbol{u}\bcdot \boldsymbol{t}_{\text{lg}}^{\text{apex}}\right)^2 \: {\rm d}A } \\
\bar{u}_{t,\text{rms}}^{\text{circum}}(t)&=\sqrt{\frac{1}{ \int_{\Gamma_{\text{lg}}} {\rm d}A} \int_{\Gamma_{\text{lg}}}  \left(\boldsymbol{u}\bcdot \boldsymbol{t}_{\text{lg}}^{\text{circum}}\right)^2 \: {\rm d}A }\,,
\label{eq:threedim:meantangvelorms}
\end{align}
where $\boldsymbol{t}_{\text{lg}}^{\text{apex}}$ is the interface tangent pointing towards the apex and $\boldsymbol{t}_{\text{lg}}^{\text{circum}}$ is the tangent pointing along the circumference of the droplet. Hence, a non-vanishing value of $\bar{u}_{t,\text{rms}}^{\text{circum}}$ with respect to $\bar{u}_{t,\text{rms}}^{\text{apex}}$ indicates the absence of axial symmetry. A typical evolution of these quantities is depicted in \figurename~\ref{fig:threedim:scheme}(d). Initially, the Marangoni instability breaks the axial symmetry, which results in an increasing value of $\bar{u}_{t,\text{rms}}^{\text{circum}}$. After that, it can take several seconds of simulation time $t_{\text{3d}}$ until the system enters its long-time behavior. The long-time behavior regime can be identified when both $\bar{u}_{t,\text{rms}}^{\text{apex}}$ and $\bar{u}_{t,\text{rms}}^{\text{circum}}$ show a converged mean value and variance. 

\begin{figure}\centering\includegraphics[width=1\textwidth]{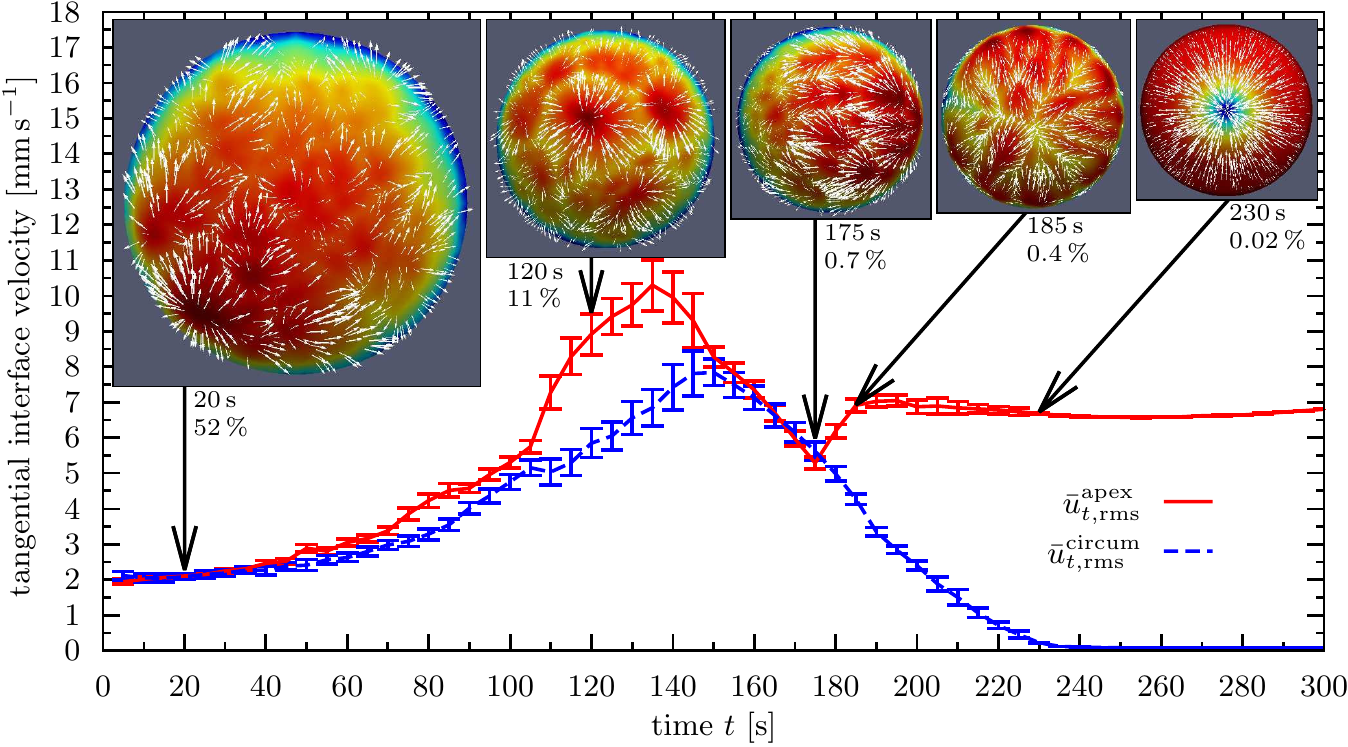}
\caption{Evolution of the rms-averaged tangential velocity components based on the long-time behavior of individual 3d-simulations corresponding to the axisymmetric simulation of \figurename~\ref{fig:axisymm:binary}. The error bars indicate the standard deviation and the insets show typical snapshots of the 3d-simulations at the indicated times. The numbers below the snapshot are the time and the averaged ethanol mass fraction.}
\label{fig:threedim:mergedevo}
\end{figure}

The characteristic flow statistics were finally extracted based on at least $t_{\text{3d}}=\SI{30}{\second}$ of long-time behavior for different snapshot times $t$. The resulting data for the tangential fluid velocity components is depicted in \figurename~\ref{fig:threedim:mergedevo}. Initially, the droplet is highly non-axisymmetric with $\bar{u}_{t,\text{rms}}^{\text{apex}}\approx{}\bar{u}_{t,\text{rms}}^{\text{circum}}$. With ongoing time, ethanol evaporates and since the composition-dependence of the surface tension has its steepest gradient in the limit of dilute ethanol, the flow becomes faster. After a maximum at about $t\approx \SI{130}{\second}$, the average velocity decreases again.
Until $t=\SI{175}{\second}$, the characteristic flow pattern is predominantly driven by solutal Marangoni flow. The ethanol concentration at the interface exhibits spatio-temporally chaotic cellular patterns. Moreover, it is possible that a net flow from one side of the rim to the opposing side can build up (e.g. from right to left in the inset at $t=\SI{175}{\second}$). After $t=\SI{175}{\second}$, i.e. after the local minimum of $\bar{u}_{t,\text{rms}}^{\text{apex}}$, a transition can be observed. The flow is mainly directed towards the apex due to the thermal Marangoni effect, but the dilute ethanol still causes symmetry-breaking perturbations (cf. inset at $t=\SI{185}{\second}$). This characteristic flow pattern has already been discussed in \figurename~\ref{fig:threedim:scheme}(c). With ongoing time and vanishing ethanol, the flow gets more and more axisymmetric and regular, which can be inferred by the vanishing value of $\bar{u}_{t,\text{rms}}^{\text{circum}}$.

\subsection{Comparison with experiment}
\label{sec:threedim:micropiv}
As described in section \ref{sec:exp:micropiv}, micro-PIV measurements were performed during the evaporation of the binary water-ethanol droplet. Thereby, experimental data for the flow in the focal plane, i.e. close to the substrate became available. Since the simulation solves the entire flow field in the bulk, it is possible to extract the corresponding velocity by slicing the three-dimensional mesh at the focal plane and projecting the local velocity vectors onto it. This allows for a direct comparison of experimentally and numerically determined velocity.

\begin{figure}
\centering
\includegraphics[width=0.92\textwidth]{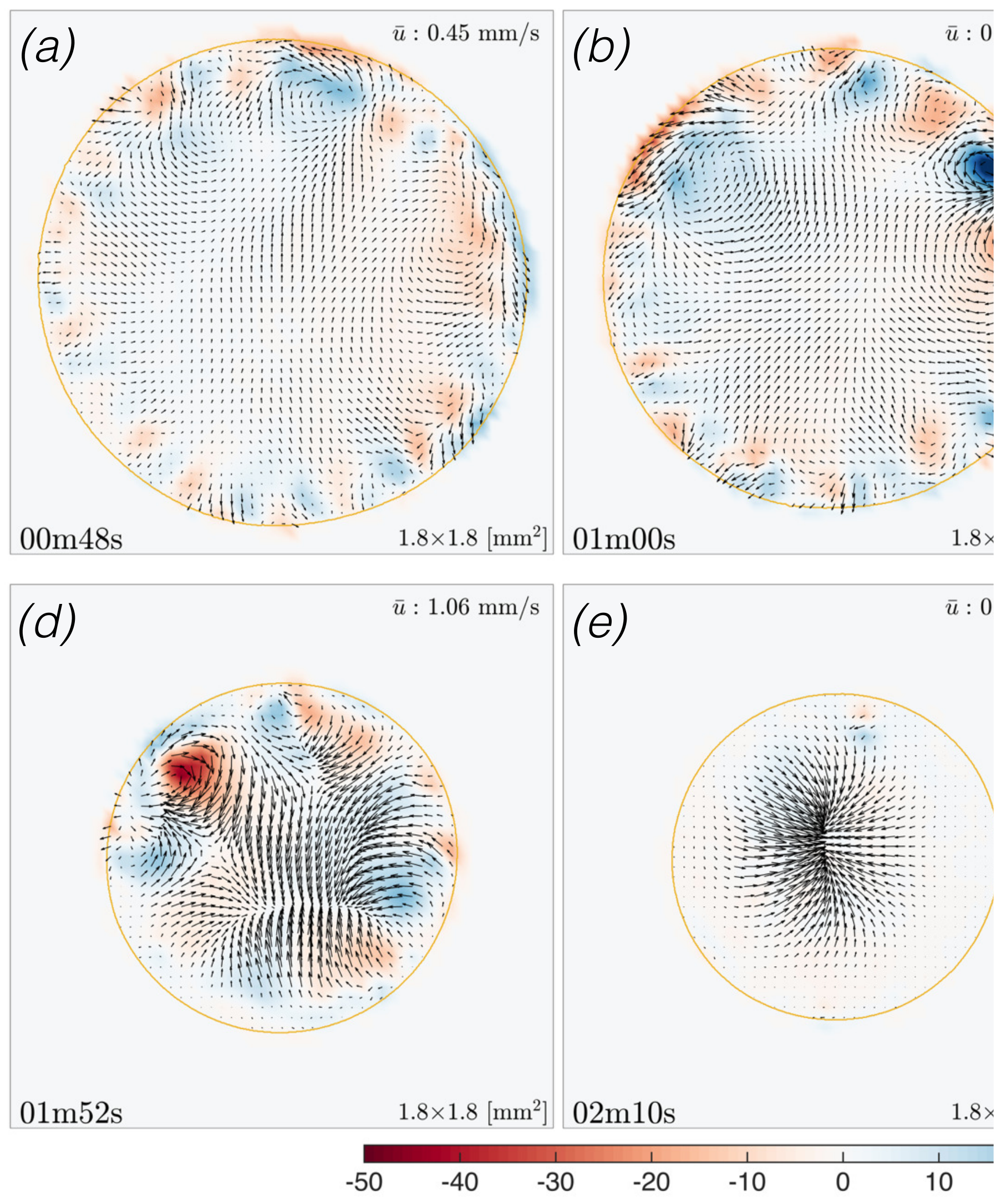}\\
\includegraphics[width=0.9125\textwidth]{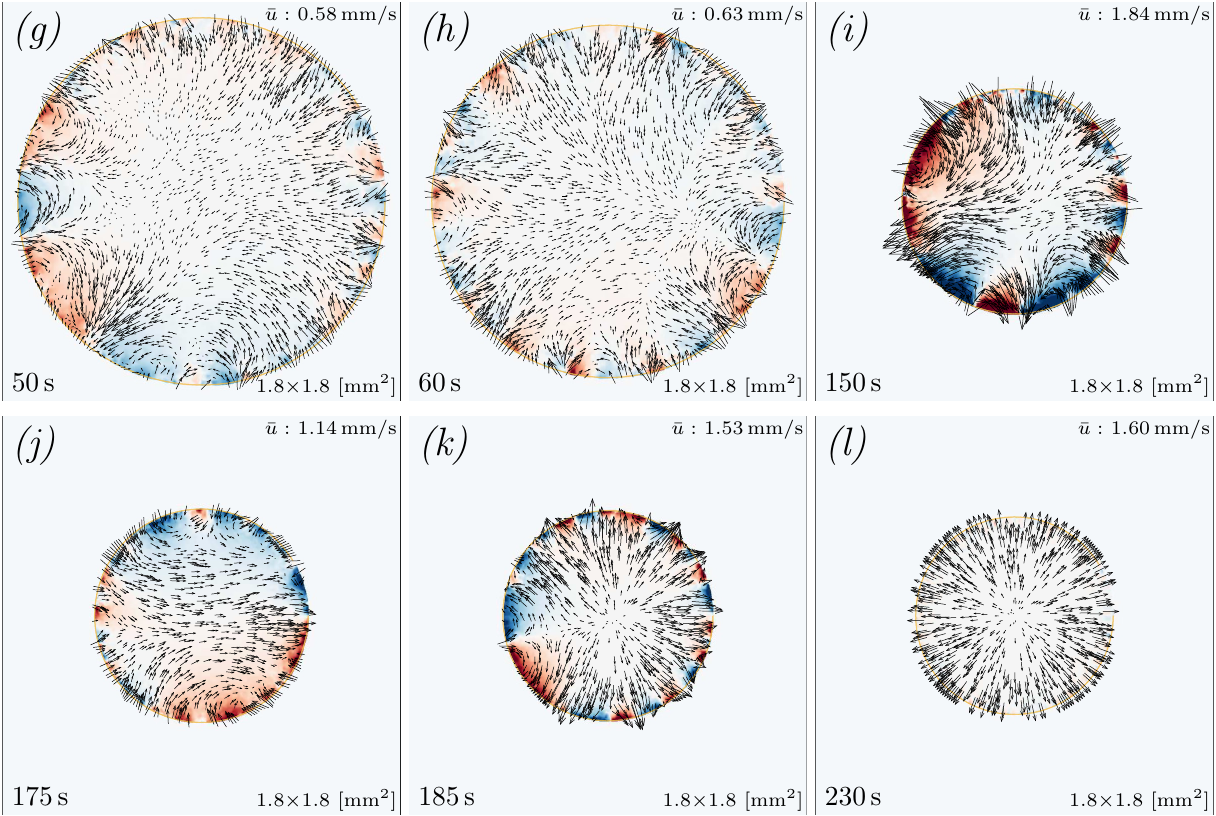}
  \caption{(a-f) Experimental snapshots of the velocity field in the focal plane at different instants. The arrows display the local velocity and the vorticity is color-coded. The yellow circle indicates the intersection of the liquid-air interface and the focal plane. (a) Small vortices are present close to the liquid-air interface, breaking the axial-symmetry in the entire droplet. (b) Small vortices can coalesce into a pairs of bigger vortices. (c,d) The overall flow velocity and vorticity increase. (e,f) After a while, the flow has an inward flow pattern (but not perfectly axisymmetric) and finally calms down. (g-l) Velocity in the focal plane as extracted from the numerics. While the simulations shows initially good agreement with the experiment, deviations are present at later times. Note that the color code is the same for both experiment (a-f) and numerics (g-l). Corresponding movies are available as supplementary material.} \label{fig:threedim:piv}  \end{figure} 

\begin{figure}\centering\includegraphics[width=1\textwidth]{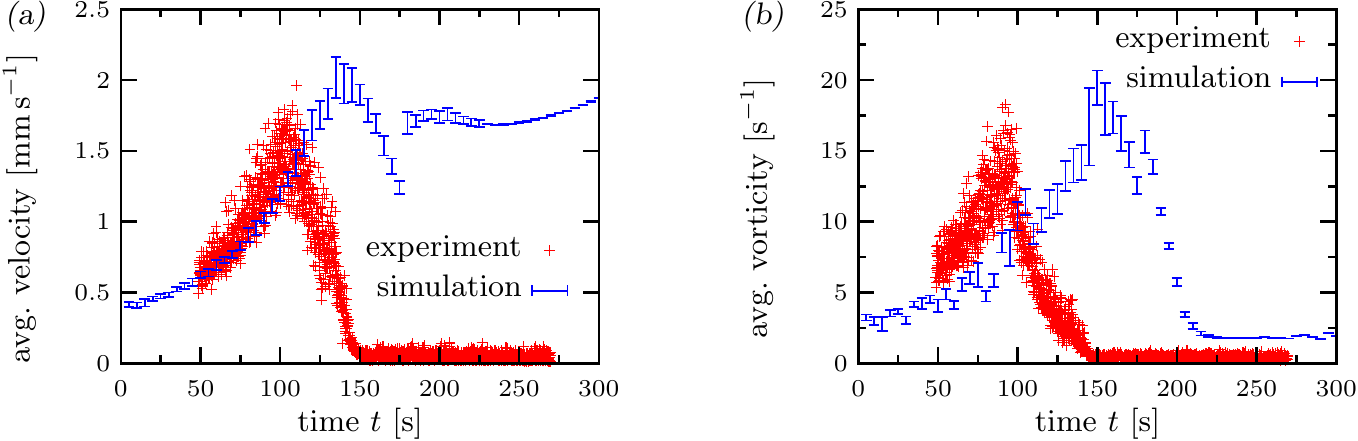}
\caption{Averaged velocity (a) and vorticity (b) in the focal plane based on the experiment and the numerical simulation. The error bars indicate the standard deviation of the numerically obtained quantities.}
\label{fig:threedim:expcmp}
\end{figure}

Representative snapshots of the experimental micro-PIV measurement are depicted in \figurename~\ref{fig:threedim:piv}(a-f). It is apparent that the axisymmetry is initially broken by the presence of multiple vortices near the liquid-air interface (a,b). With ongoing time, the flow gets more intense and more chaotic up to a maximum at about $t=\SI{100}{\second}$ (c,d). Directly after that, a net flow towards the center of the focal plane is building up, which is not perfectly axisymmetric but globally it is directed radially inward (e). The flow is slowing down until it completely stops at about $t=\SI{140}{\second}$ (f).

The data extracted from the numerical simulation is depicted in \figurename~\ref{fig:threedim:piv}(g-l) for comparison. It is apparent how the typical flow is very similar at early times (g,h), but in the numerics the maximum averaged flow intensity can only be found at $t\approx\SI{150}{\second}$ (i,j), i.e. when the flow has already ceased in the experiments. After this maximum, the flow does not turn inward, but outward instead, which is caused by the thermal Marangoni effect. As long ethanol resides in the droplet, deviations from perfect axisymmetric thermal Marangoni flow can be seen (k). These eventually decrease over time until the flow finally converges to a perfectly axisymmetric intense thermal Marangoni flow.
The initial agreement of experiment and simulation can also be inferred from \figurename~\ref{fig:threedim:expcmp}(a), where the averaged velocities in the focal plane are plotted following both methods. At later times, however, the disagreement is obvious.

Since the disagreement is most pronounced at later times, it cannot be attributed to the presence of a binary mixture. In fact, for $t\gtrsim\SI{250}{\second}$ the droplet is virtually a pure water droplet, which should show an intense thermal Marangoni flow according to the simulation, which is not observed in experiment. This disagreement between prediction and observation of thermal Marangoni flow in water droplets is well-known \citep{Hu2005b}. At $t=\SI{250}{\second}$, the numerically obtained temperature difference between contact line and apex reads $\Delta T\approx\SI{0.57}{\kelvin}$, which corresponds to a surface tension difference of $\Delta \sigma\approx-\SI{0.084}{\milli\newton\per\meter}$, i.e. $\SI{0.1}{\percent}$ of the averaged surface tension. As already discussed by \citet{Hu2005b}, this small thermally induced difference can be counteracted by a very small amount of surface-active contaminants, which are unavoidable during the experiment. 

Therefore, the present scenario, i.e. good agreement at initial stages and large disagreement in the limit of pure water, can presumably be attributed to the presence of contaminants. Since the initial solutal Marangoni instability is intense and chaotic, the contaminants are unable to arrange to a counteracting distribution along the interface. With ongoing time, three mechanisms can contribute to an increase of the overall contaminant concentration at the interface. These are the fact that the droplet and thus the interface are shrinking, a possible ongoing adsorption of contaminants from the gas phase during the evaporation process, and a possibly increasing interface affinity of the contaminants with decreasing ethanol concentration. Once the contaminant concentration is sufficiently high, they start to stabilize the solutal Marangoni flow leading to the almost symmetric radially inward flow as seen in \figurename~\ref{fig:threedim:piv}(e). Eventually, the thermal Marangoni flow is completely suppressed by the contaminants (cf. \figurename~\ref{fig:threedim:piv}(f)). However, to substantiate this scenario, more elaborate experiments, in particular measurements of the contaminant concentration at the interface, or a generalization of the model considering surface-active contaminants would be necessary. It can also not be ruled out that the hydrophobic polystyrene tracer particles influence the interface dynamics. 

\section{A Closer Look at the Oil-Microdroplets via Confocal Microscopy}
\label{sec:confocal:confocal}

\begin{figure}\centering\includegraphics[width=1\textwidth]{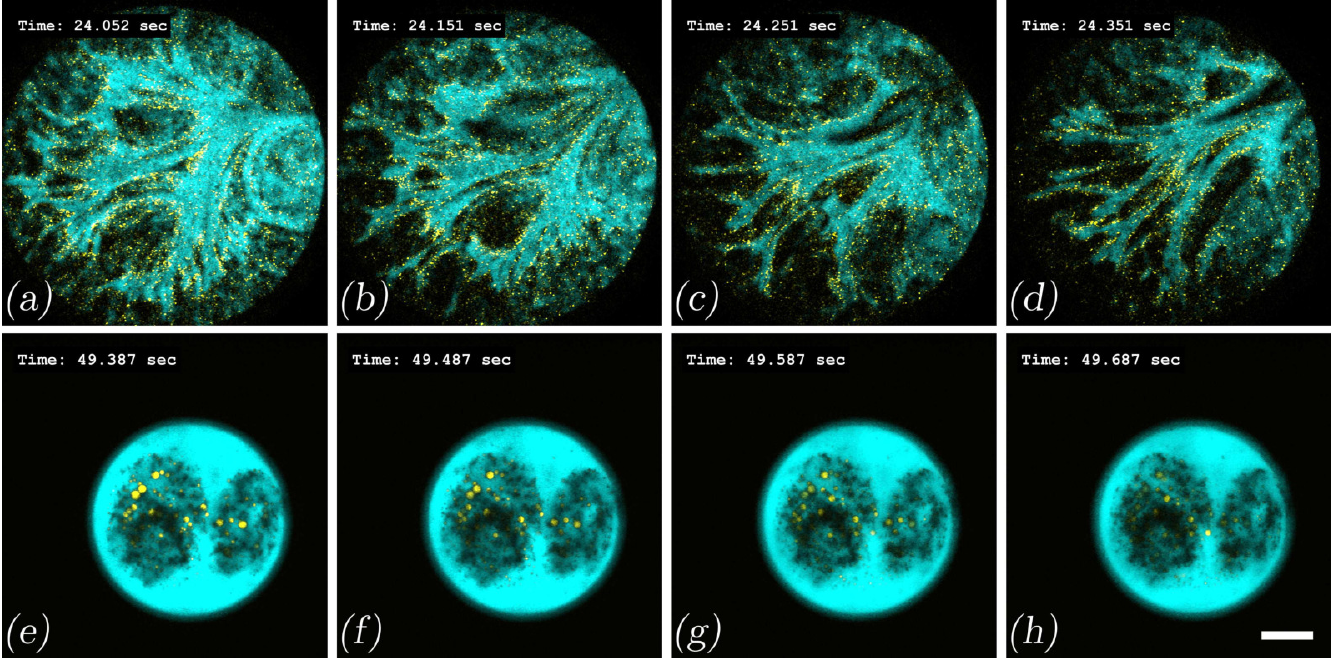}
\caption{The complicated flow inside the Ouzo droplet. The black areas inside the droplet represent the shadow of the oil-microdroplets presenting in between the focal plane and the objective. The scale bar is $\SI{100}{\micro\meter}$. A corresponding movie is available as supplementary material.}
\label{fig:confocal:flow}
\end{figure}

\begin{figure}\centering\includegraphics[width=1\textwidth]{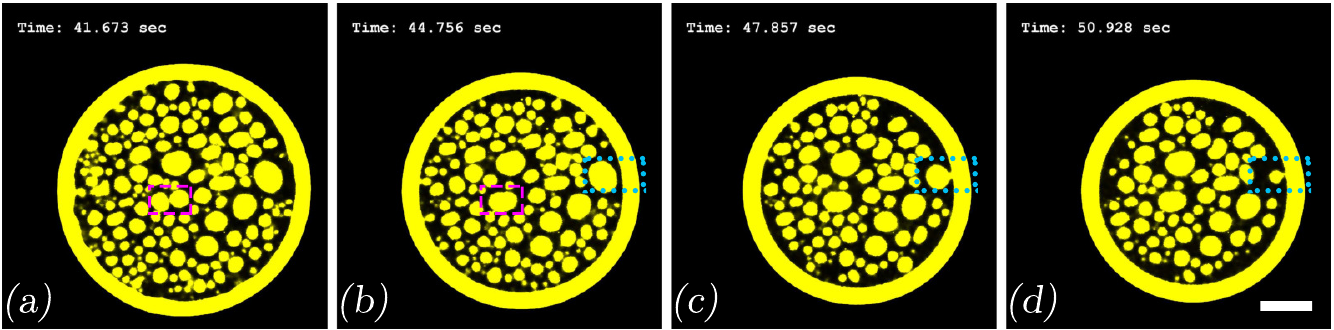}
\caption{Coalescence of oil-microdroplets on the substrate forming the oil ring. Only the images of the oil phase are presented. Oil-microdroplets within the purple dashed rectangle in (a) coalesce to form a merged surface microdroplet (see the corresponding position in (b)). The oil-microdroplet within the blue dotted rectangle in (b), (c), and (d) is gradually absorbed by the shrinking oil ring. The scale bar is $\SI{100}{\micro\meter}$. A corresponding movie is available as supplementary material.}
\label{fig:confocal:ring}
\end{figure}

While the presence of the oil-microdroplets in the Ouzo droplets does not allow for micro-PIV measurements, confocal microscopy, as described in section \ref{sec:exp:confocal}, represents the ideal technique to monitor the flow as well as the behavior of the oil-microdroplets inside an evaporating Ouzo droplet. 

In \figurename~\ref{fig:confocal:flow}, the complicated flow inside the Ouzo droplet was captured by focusing on the position just above the oil ring. Here, the trans-anethole oil and the water-ethanol mixture are colored in yellow and blue, respectively. The upper panel (i.e., (a)-(d)) shows the initial regime of the evaporation when the Ouzo effect started. A strong chaotic solutal Marangoni flow can be seen, revealing the axial symmetry breaking. The lower panel (i.e., (e)-(h)) exhibits the following onset of the flow transition, leading to the stop of the convection flow when ethanol was almost gone. A huge contrast in the flow pattern before and after the flow transition is revealed, which is in particular apparent from the corresponding movie provided as supplementary material. The flow transition in the Ouzo droplet is qualitatively the same as in the binary water-ethanol droplet determined by the micro-PIV experiment (cf. section \ref{sec:threedim:micropiv}).

In \figurename~\ref{fig:confocal:ring}, the behavior of the oil-microdroplets on the substrate is revealed by focusing on the bottom of the Ouzo droplet. The growth of these oil-microdroplets is demonstrated to be predominantly induced by coalescence rather than by Ostwald ripening, which is usually thought to occur in the bulk of a phase-separating Ouzo mixture \citep{Sitnikova2005a}. However, a minor contribution of Ostwald ripening in the Ouzo droplet cannot be ruled out. The oil ring is almost entirely emerging by coalescence of oil droplets near the contact line.
\section{Conclusion}
\label{sec:conclusion:conclusion}
In conclusion, we have investigated the evaporation of multi-component droplets with numerical and experimental methods. By comparing the results of these methods, we are able to draw the following main conclusions:

Due to the good agreement of the numerical predictions and the experimental data for a pure water droplet and a binary water-ethanol droplet, we have validated the axisymmetric finite element method model of \citet{Diddens2017b}. It has been shown that the quality of the agreement decisively depends on the consideration of the interplay of multi-component evaporation and thermal effects. In particular, it has been shown that even for a pure water droplet the accuracy of the isothermal model of \citet{Popov2005a} is limited if the substrate is thin. Any noticeable influence of Marangoni flow and Stefan flow in the gas phase on the evaporation has been ruled out for these droplets. While the pure water droplet and the binary water-ethanol droplet are in perfect agreement with the experiment, the simulation of the ternary Ouzo droplet initially shows good agreement, including the onset of oil nucleation, but exhibits a faster evaporation of the remaining water residual than in experiment. This issue can presumably be attributed to the presence of the oil ring, i.e. a geometric deviation from the typical spherical cap shape, and which was not included in the present model.

To experimentally investigate the flow in the binary water-ethanol droplet, micro-PIV measurements have been performed. Since the flow is clearly non-axisymmetric as long as ethanol is present, the model had to be generalized to a full three-dimensional version. While the data is initially in good quantitative agreement, deviations between experiment and simulation can be found at later times. In particular, the simulation shows an intense thermal Marangoni flow once pure water remains, whereas the micro-PIV measurement shows no flow at all. Since all other aspects, i.e. composition-dependent liquid properties and thermal influences, are considered in the model, the mismatch can only be attributed to the presence of surface-active contaminations. This observation encourages the development of even more elaborate models which will take the role of surfactants into account and the conduction of novel experiments which try to completely exclude any contamination.

Although we were unable to perform micro-PIV measurements in the Ouzo droplet, the flow inside this droplet was qualitatively revealed by the use of confocal microscopy. By visualizing the trajectories of the oil microdroplets, similar flow transitions as in the micro-PIV results of the binary water-ethanol droplet were found. Furthermore, we found that the oil ring at the rim of the Ouzo droplet predominantly emerges due to coalescence of droplets, which is different from the Ouzo effect in the bulk, which is thought to be primarily constituted by Ostwald ripening. 

\begin{acknowledgments}
The authors thank Chao Sun, Varghese Mathai and Alvaro Marin for the useful discussion on the micro-PIV technique.
C.D. and J.G.M.K. gratefully acknowledge funding by the Dutch Technology Foundation STW, which is part of the Netherlands Organization for Scientific Research (NWO) and partly funded by the Ministry of Economic Affairs. H.T. thanks for the financial support from the China Scholarship Council (CSC, file No. 201406890017). We also acknowledge the Dutch Organization for Research (NWO) and the Netherlands Center for Multiscale Catalytic Energy Conversion (MCEC) and the Max Planck Center Twente for Complex Fluid Dynamics for financial support. This work is part of an Industrial Partnership Programme (IPP) of the Foundation for Fundamental Research on Matter (FOM), which is financially supported by the Netherlands Organization for Scientific Research (NWO). This research programme is co-financed by Oc\'{e}-Technologies B.V., Twente University and Eindhoven University.
\end{acknowledgments}

\bibliographystyle{jfm}
\bibliography{refs}

\end{document}